\documentclass{article}
\usepackage[final]{neurips_bdl2019}

\usepackage{amssymb}
\usepackage{mathrsfs}
\usepackage{booktabs}
\usepackage{xcolor}
\usepackage{xspace}
\usepackage{hyperref}

\usepackage{xargs}
\usepackage{bm}
\usepackage{graphicx}
\usepackage{booktabs}

\usepackage{pbox}

\usepackage{wrapfig}
\usepackage{pgfplots}

\usepackage{amsmath}

\graphicspath{{./figs/}}
\usepackage{algorithm}
\usepackage{algpseudocode}
\newfloat{algorithm}{t}{lop}

\usepackage[noabbrev,capitalise,nameinlink]{cleveref}

\usepackage{wrapfig}

\usepackage{tikz}

\makeatletter
\let\MYcaption\@makecaption
\makeatother

\usepackage{caption}
\usepackage{subcaption}
\usepackage{tikz}

\usepackage[colorinlistoftodos,prependcaption,textsize=small]{todonotes}

\usepackage{etoolbox}

\usepackage{dblfloatfix}

\newcommandx{\david}[2][1=]%
{\todo[disable,#1]{#2}}
\newcommandx{\kathrin}[2][1=]%
{\todo[disable,#1]{#2}}
\newcommandx{\td}[2][1=]{\todo[inline,linecolor=blue,backgroundcolor=blue!25,bordercolor=blue,#1]{\textbf{Todo}: #2}}

\newcommand{\hcluex}{HCLU examples}
\newcommand{\hclu}{HCLU adversarial examples}

\author{Kathrin Grosse \\
  CISPA Helmholtz Center\\
  \texttt{kathrin.grosse@cispa.saarland} \\
  \And
  David Pfaff \\
  CISPA Helmholtz Center\\
  david.pfaff@cispa.saarland
  \And
  Michael T. Smith \\
  Sheffield University \\
  m.t.smith@sheffield.ac.uk \\
  \And
  Michael Backes \\
  CISPA Helmholtz Center \\
  backes@cispa.saarland}

\begin{document}

\title{The Limitations of Model Uncertainty in Adversarial Settings}

\maketitle

\begin{abstract}
Machine learning models are vulnerable to adversarial examples: minor perturbations to input samples intended to deliberately cause misclassification. While an obvious security threat, adversarial examples 
yield as well insights about the applied model itself. 
We investigate adversarial examples in the context of Bayesian neural network's (BNN's) uncertainty measures. 
As these measures are highly non-smooth, we 
use a smooth Gaussian process classifier (GPC) as substitute. 
We show that both confidence and uncertainty can be unsuspicious even if the output is wrong. 
Intriguingly, we find subtle differences in the features influencing uncertainty and confidence for most tasks. 
\end{abstract}


\section{Introduction}
\label{sec:intro}
Machine learning classifiers are used for various purposes in a variety of research and industry applications. 
However, these classifiers have been shown to be vulnerable to a number of different attacks~\citep{%
DBLP:conf/ceas/LowdM05,mei2015using,goodfellow2015explaining,biggio2018wild}.
\emph{Adversarial examples}, or evasion attacks, present a direct threat to classification at test-time. 
The classifier outputs a wrong label given an input sample with original utility, however slightly modified by the attacker.
In the area of computer vision, adversarial examples are  often visually indistinguishable images which are are misclassified by state-of-the-art models~\citep{goodfellow2015explaining,Moosavi-Dezfooli_2016_CVPR}. 

Adversarial examples and Bayesian uncertainty have been investigated before. \cite{DBLP:journals/corr/abs-1811-12335} show the importance of priors in robustness. 
 \cite{DBLP:journals/corr/abs-1806-00667} propose an attack to sample garbage examples in the pockets of the uncertainty of BNN. 
BNN are further investigated by \cite{DBLP:conf/icml/LouizosW17}, \cite{2017arXiv171108244R} and \cite{liu2018advbnn}. All these authors test simple fast-gradient-sign adversarial examples on (variants of) BNN and find notable differences in model uncertainty for adversarial examples. Further, \cite{DBLP:journals/corr/LiG17b} observe differences for high confidence adversarial examples. 
We show that adversarial examples exist that show no deviation for uncertainty measures, are visually similar to the original, and yet are misclassified by BNNs.

\section{Experimental Setup and Results}\label{sec:exp_study}
We extend the notion of adversarial examples from a classifiers' output to Bayesian confidence and uncertainty. Here, we assume a binary classifier. Confidence is a real value between 0 (not at all confident) and 1 (very confident). Further, uncertainty is a real larger than 0, where larger means more uncertain. We formalize the resulting optimization problem of high-confidence-low-uncertainty (HCLU) examples:
\begin{equation*}
\begin{aligned}
& \underset{\delta}{\text{min}}
& & \parallel \delta \parallel_2 \\
& \text{s.t.}
& &  \text{confidence}(f(x+\delta))>0.95, \\
& \text{and} && \text{uncertainty}(f(x+\delta)) \leq \text{uncertainty}(f(x)),
\end{aligned}
\end{equation*}
where we minimize the perturbation $\delta$ using the $L_2$ norm. An extension to other norms (as by \cite{DBLP:journals/corr/CarliniW16a}), or a restriction to change only a subset of the features, is straight forward. In general, we minimize the change between benign sample $x$ and its corresponding adversarial example $x+\delta $. The first constraint maximize the confidence of classifier $f$ and yields misclassification (e.g., the initial confidence was small). The second constraint limits the uncertainty.

Yet, as \cite{smith2018understanding} show in their work, the uncertainty estimates of BNNs are highly non-smooth, and the above problem potentially hard to solve. We 
thus decide to use a GPC as a substitute, since its decision surface is smoother. 

\textbf{Setup.} We use spam~\citep{Lichman:2013} 
data and two sub-tasks of each  MNIST~\citep{lecun-98} and Fashion MNIST~\citep{xiao2017fashion}. On Fashion MNIST, we train on trousers vs. ankle boots (one vs. nine) and sandals vs. sneakers (five vs. seven). On MNIST, we train on the one vs. nine and three vs. eight sub-tasks. We choose small tasks to fit scalability of the GPC substitute, which is further optimized on only 750 samples. Our experiments are implemented in Python. 
For deep neural networks (DNN) and BNN, we use Tensorflow~\citep{DBLP:conf/osdi/AbadiBCCDDDGIIK16},  and GPy~\citep{lawrence2004gaussian} for GPC.
The attack on GPC is implemented using the optimization routines of SciPy~\citep{scipy} and L-BFGS-B~\citep{zhu1997algorithm}. Our attack is publicly available in the adversarial robustness toolbox~\citep{art2018}, which we also used for \cite{DBLP:journals/corr/CarliniW16a}'s $L_2$ attack.


\begin{figure}
\begin{subfigure}[b]{0.105\linewidth}

\subcaption*{\textbf{1}: HC vs. \newline benign \newline }
\end{subfigure}
\begin{subfigure}[b]{0.105\linewidth}
\includegraphics[width=\linewidth]{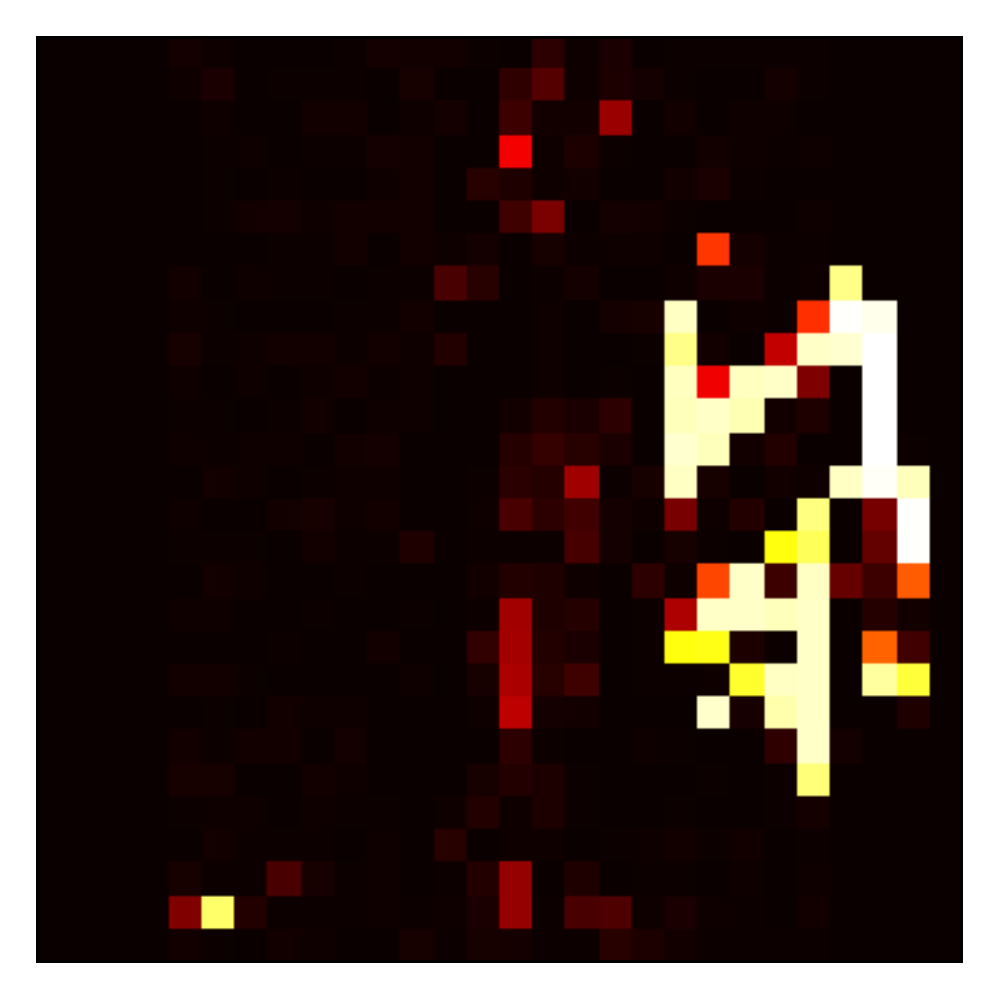}
\end{subfigure}
\begin{subfigure}[b]{0.105\linewidth}
\includegraphics[width=\linewidth]{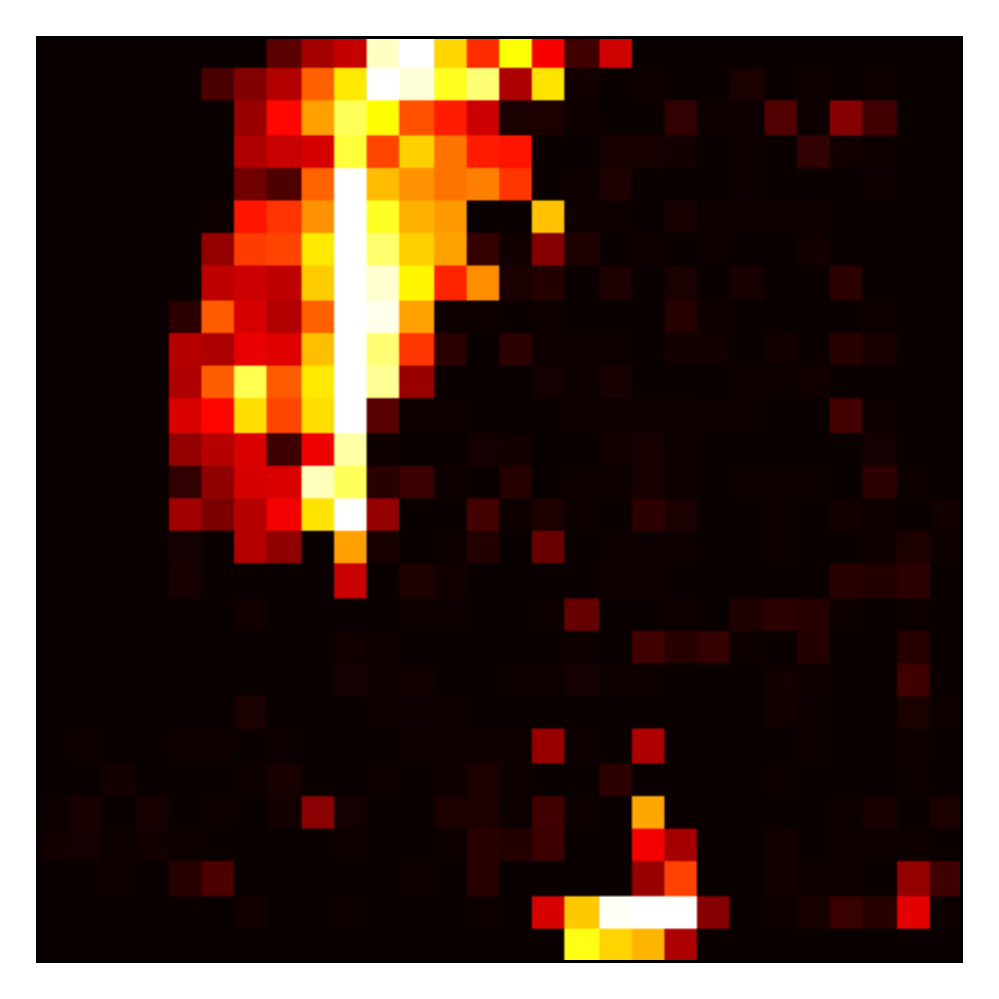}
\end{subfigure}
\begin{subfigure}[b]{0.105\linewidth}
\includegraphics[width=\linewidth]{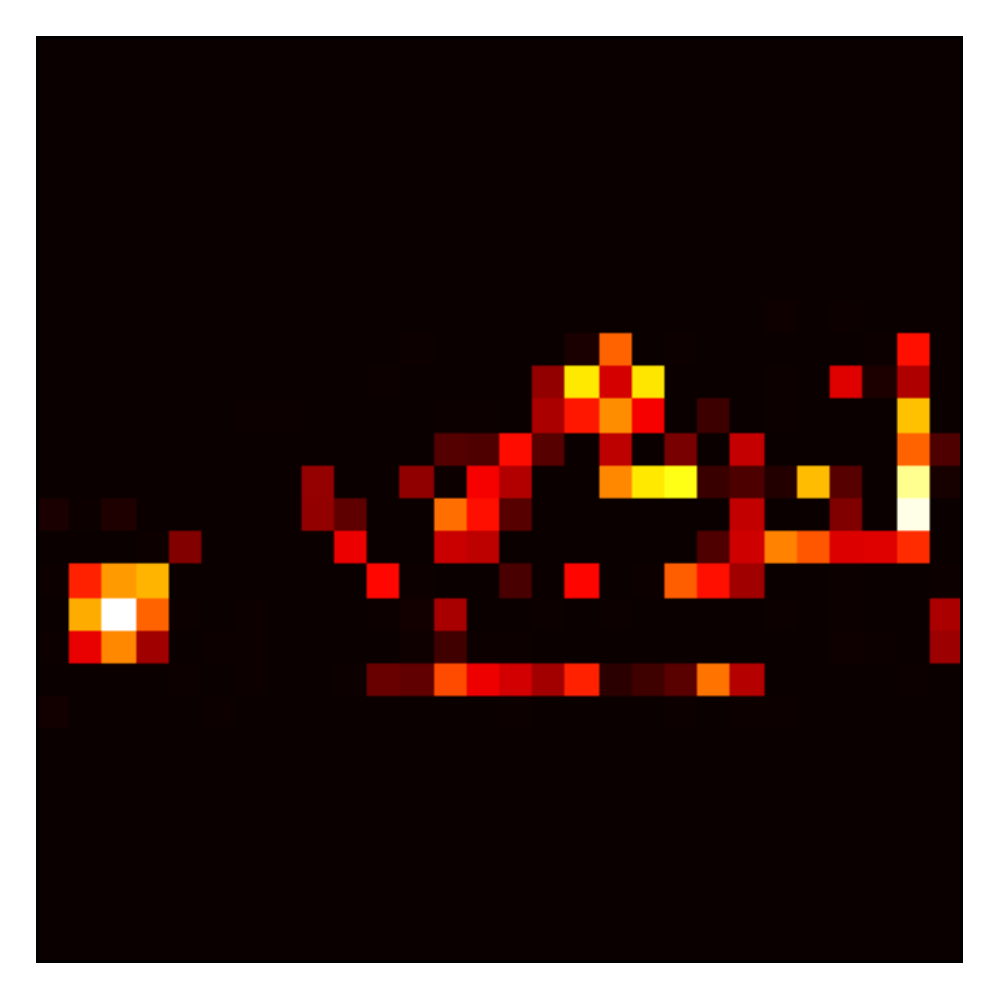}
\end{subfigure}
\begin{subfigure}[b]{0.105\linewidth}
\includegraphics[width=\linewidth]{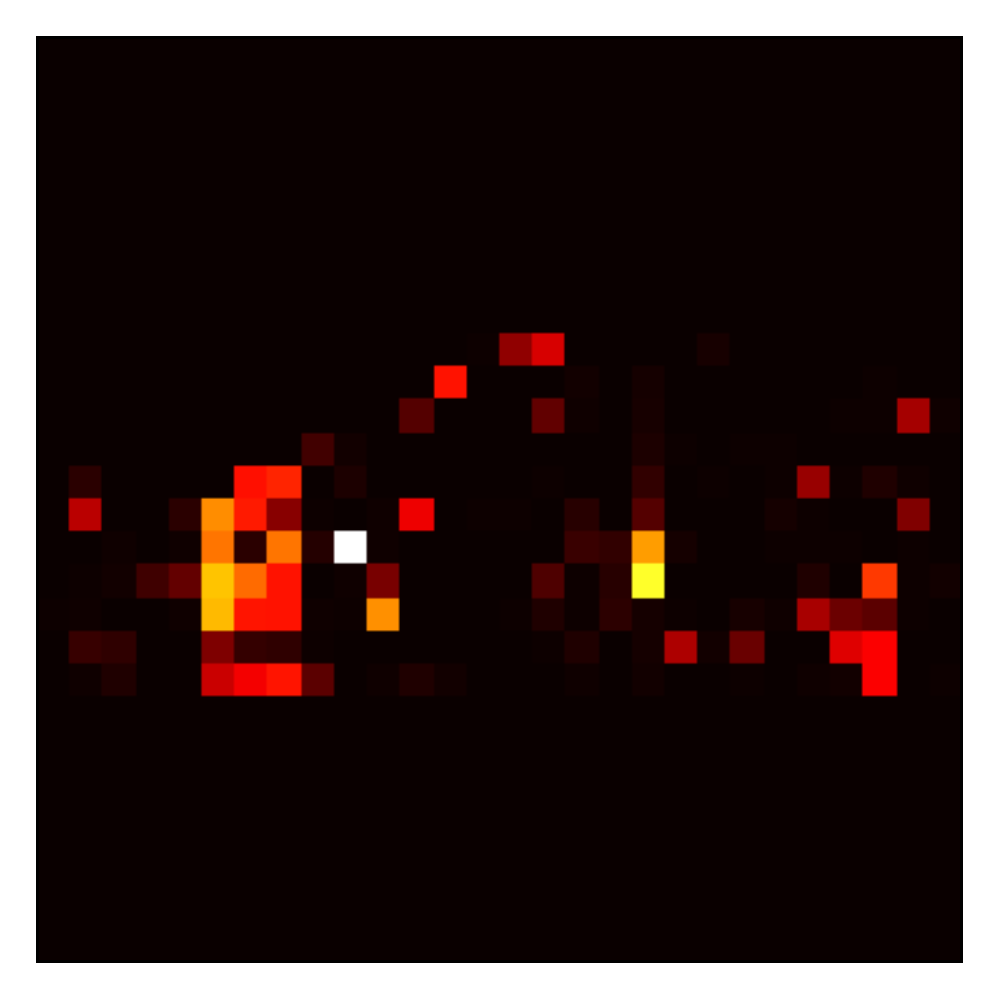}
\end{subfigure}
\begin{subfigure}[b]{0.105\linewidth}
\includegraphics[width=\linewidth]{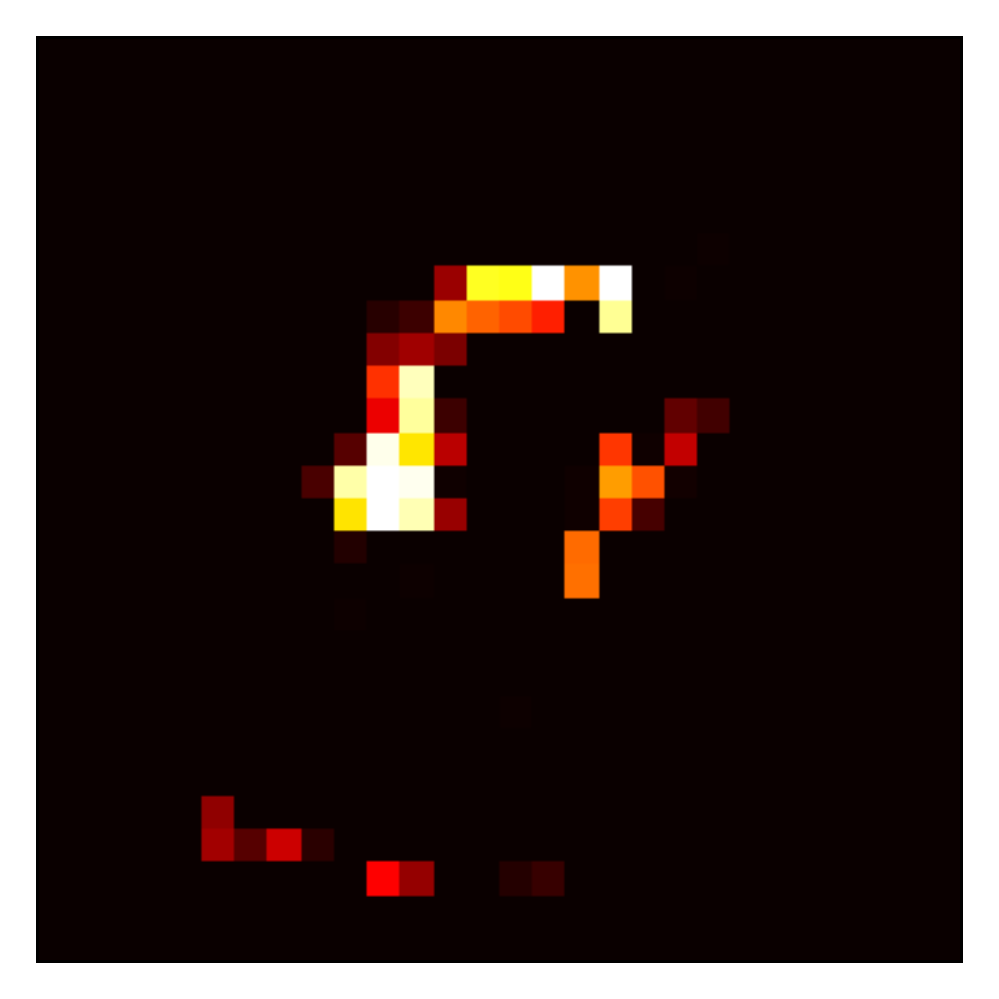}
\end{subfigure}
\begin{subfigure}[b]{0.105\linewidth}
\includegraphics[width=\linewidth]{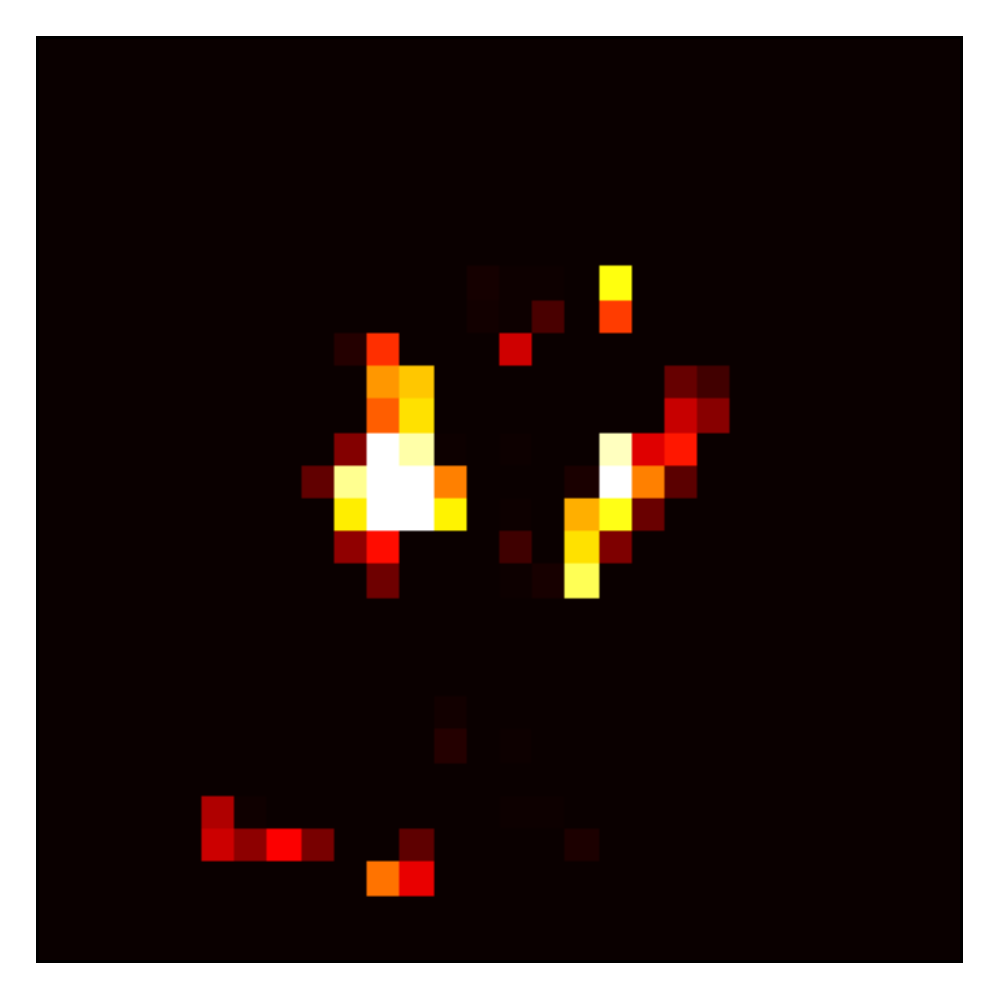}
\end{subfigure}
\begin{subfigure}[b]{0.105\linewidth}
\includegraphics[width=\linewidth]{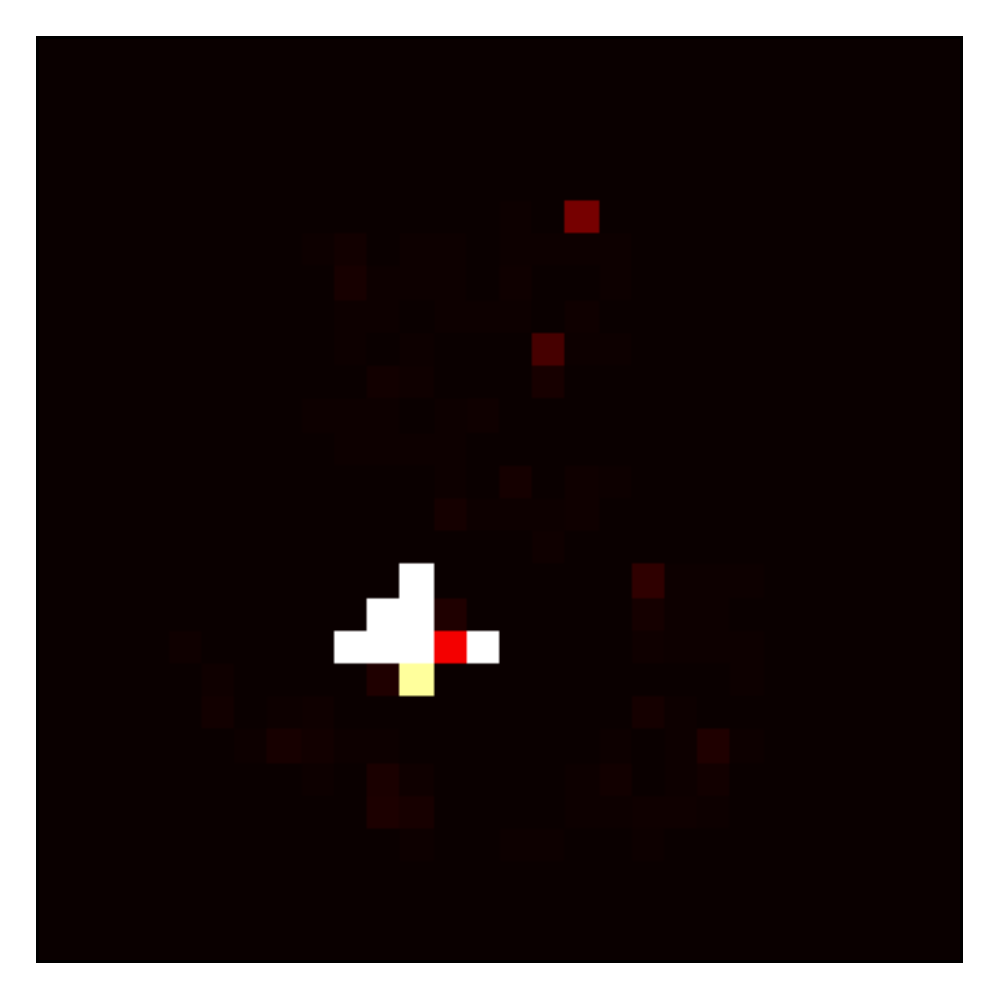}
\end{subfigure}
\begin{subfigure}[b]{0.105\linewidth}
\includegraphics[width=\linewidth]{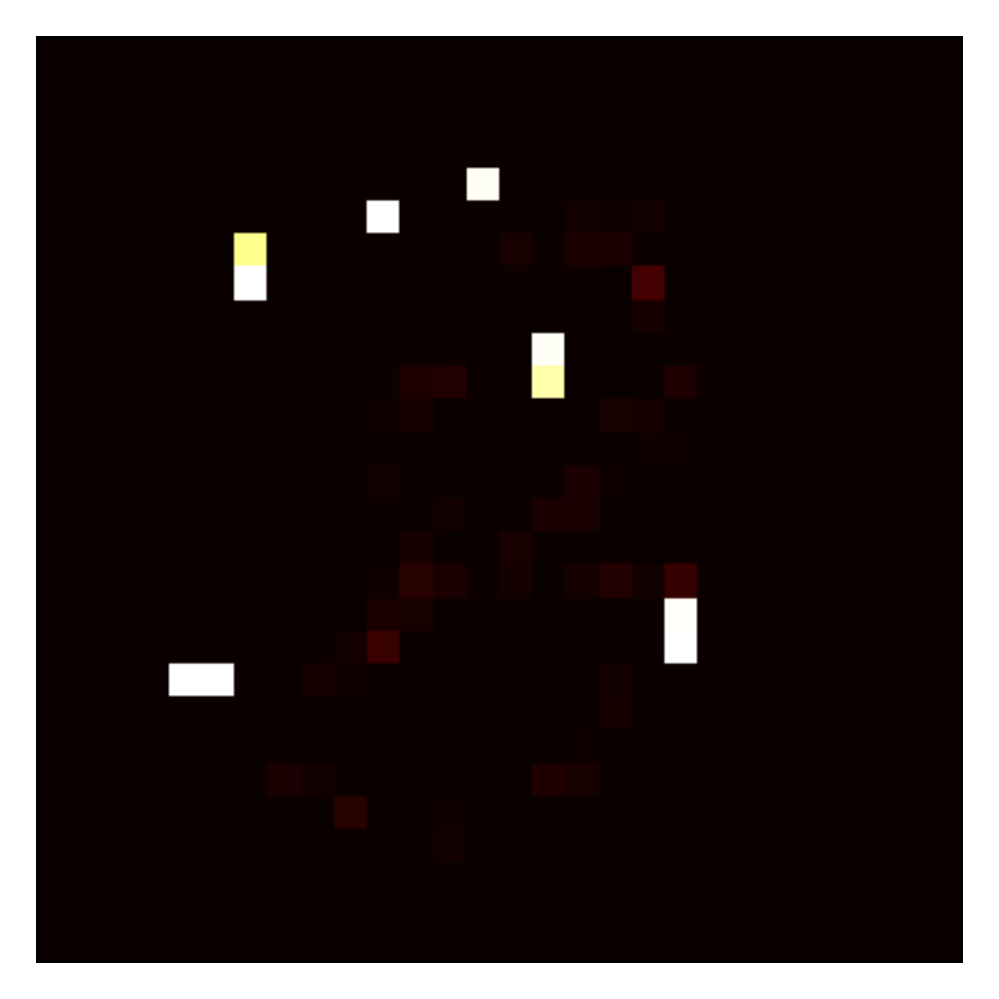}
\end{subfigure}

\begin{subfigure}[b]{0.105\linewidth}

\subcaption*{\textbf{2}: Benign \newline original \newline }
\end{subfigure}
\begin{subfigure}[b]{0.105\linewidth}
\includegraphics[width=\linewidth]{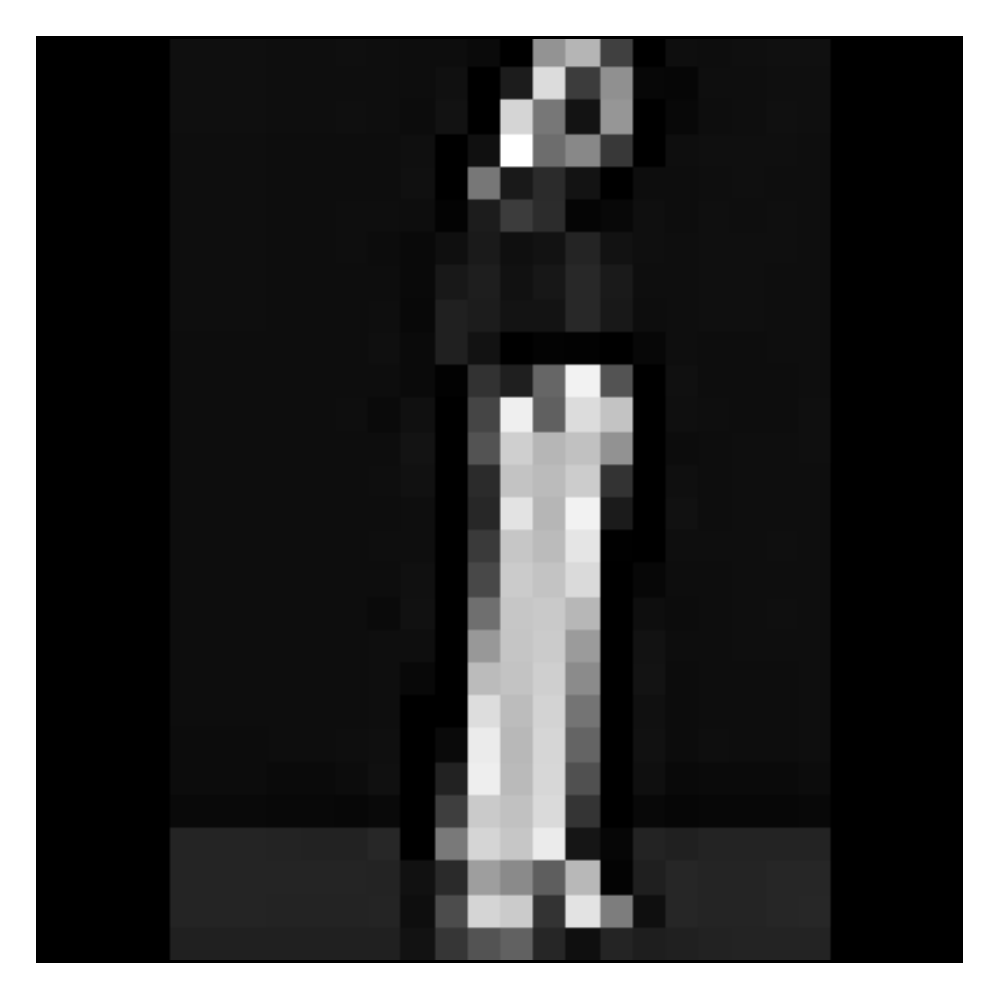}

\end{subfigure}
\begin{subfigure}[b]{0.105\linewidth}
\includegraphics[width=\linewidth]{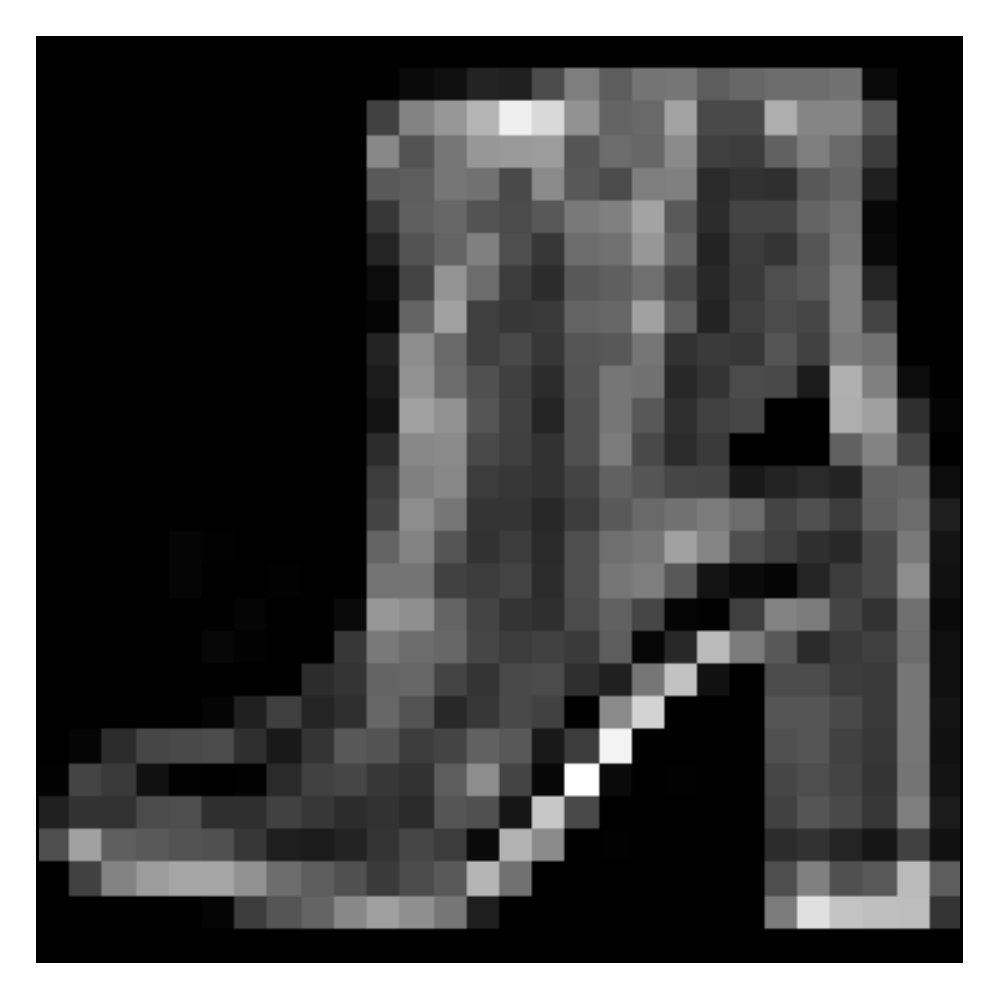}

\end{subfigure}
\begin{subfigure}[b]{0.105\linewidth}
\includegraphics[width=\linewidth]{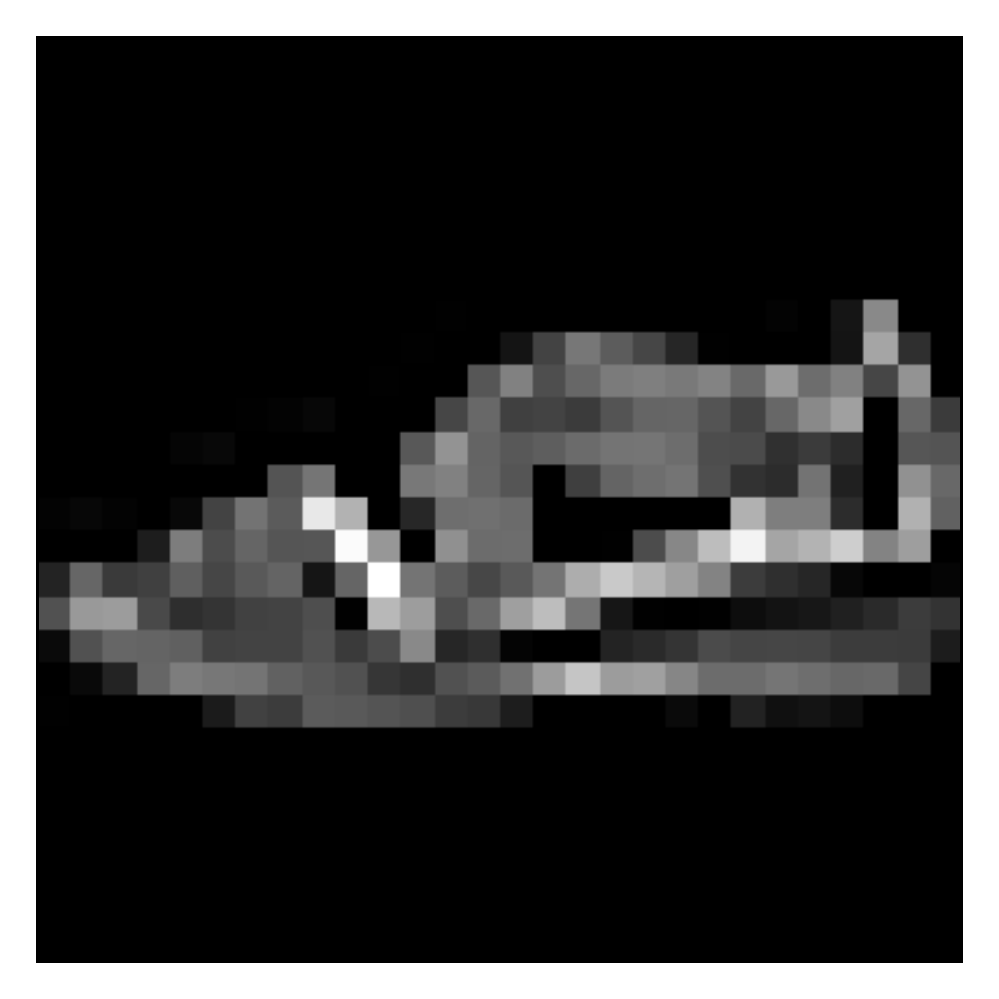}

\end{subfigure}
\begin{subfigure}[b]{0.105\linewidth}
\includegraphics[width=\linewidth]{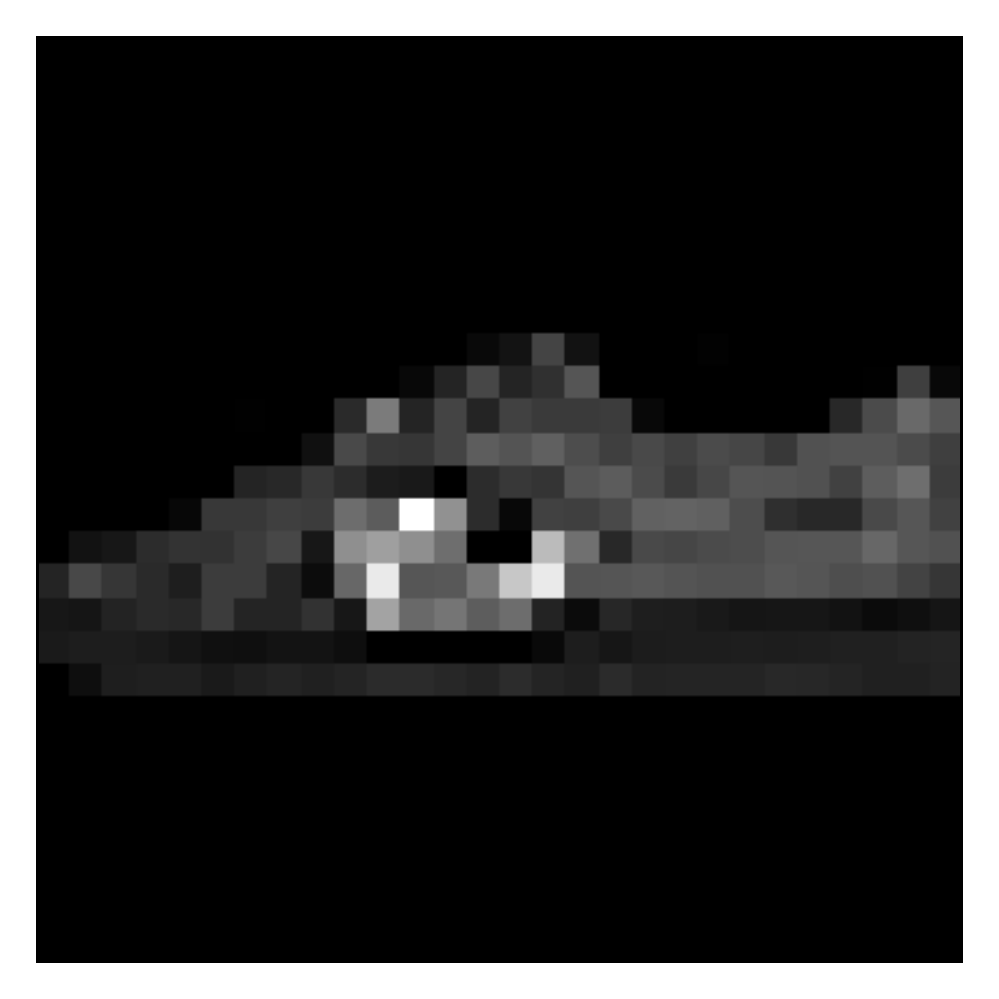}

\end{subfigure}
\begin{subfigure}[b]{0.105\linewidth}
\includegraphics[width=\linewidth]{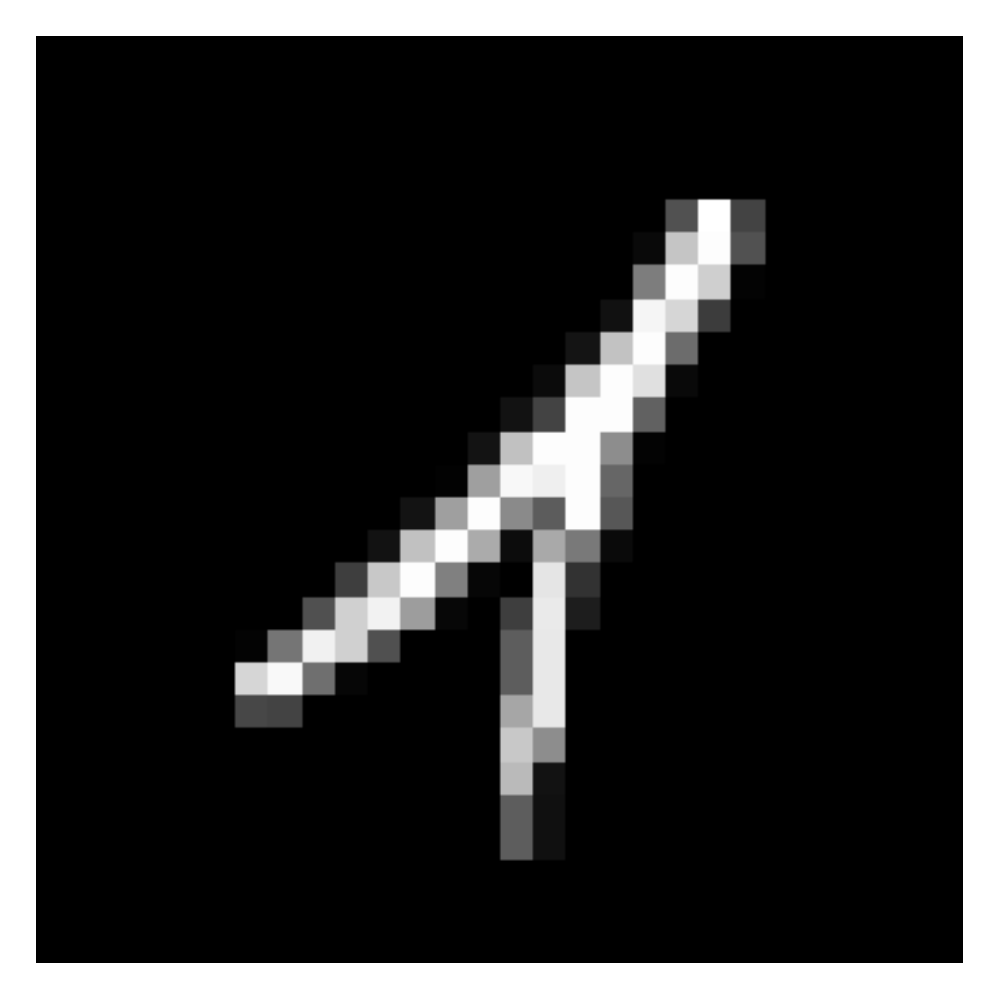}

\end{subfigure}
\begin{subfigure}[b]{0.105\linewidth}
\includegraphics[width=\linewidth]{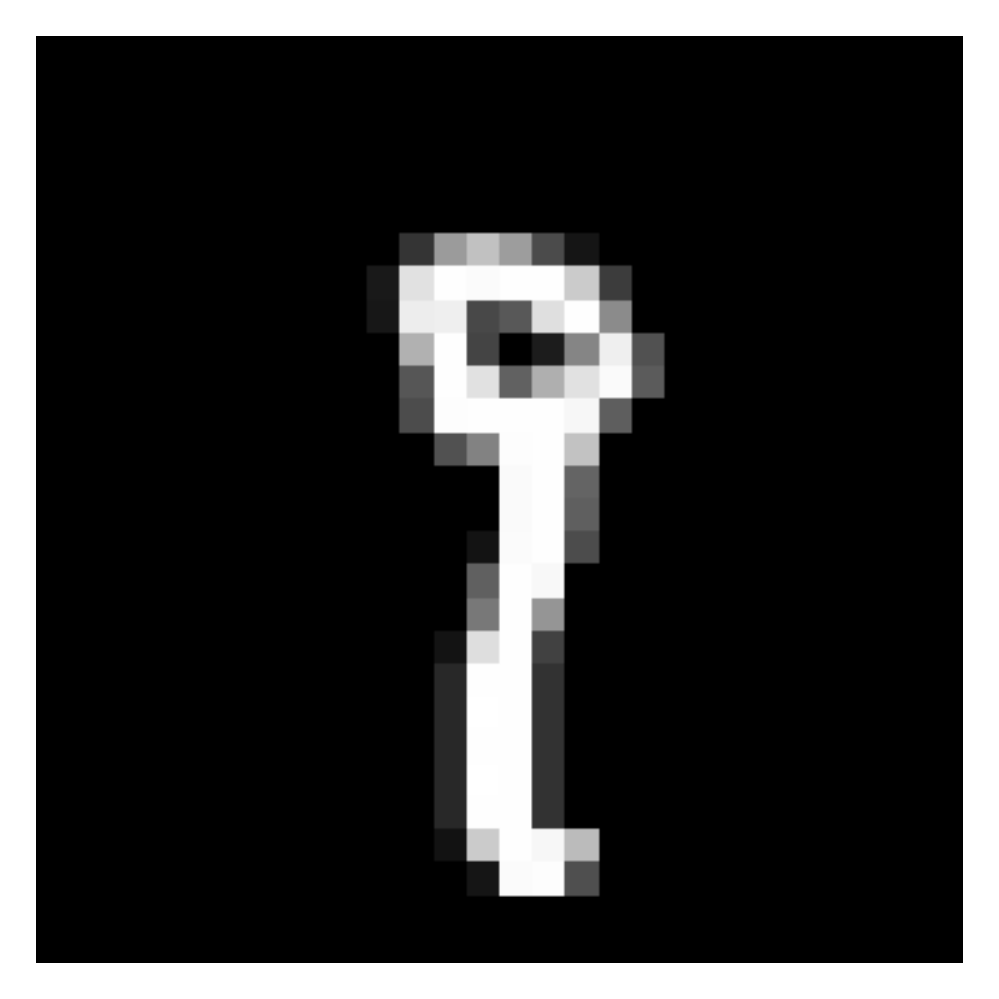}

\end{subfigure}
\begin{subfigure}[b]{0.105\linewidth}
\includegraphics[width=\linewidth]{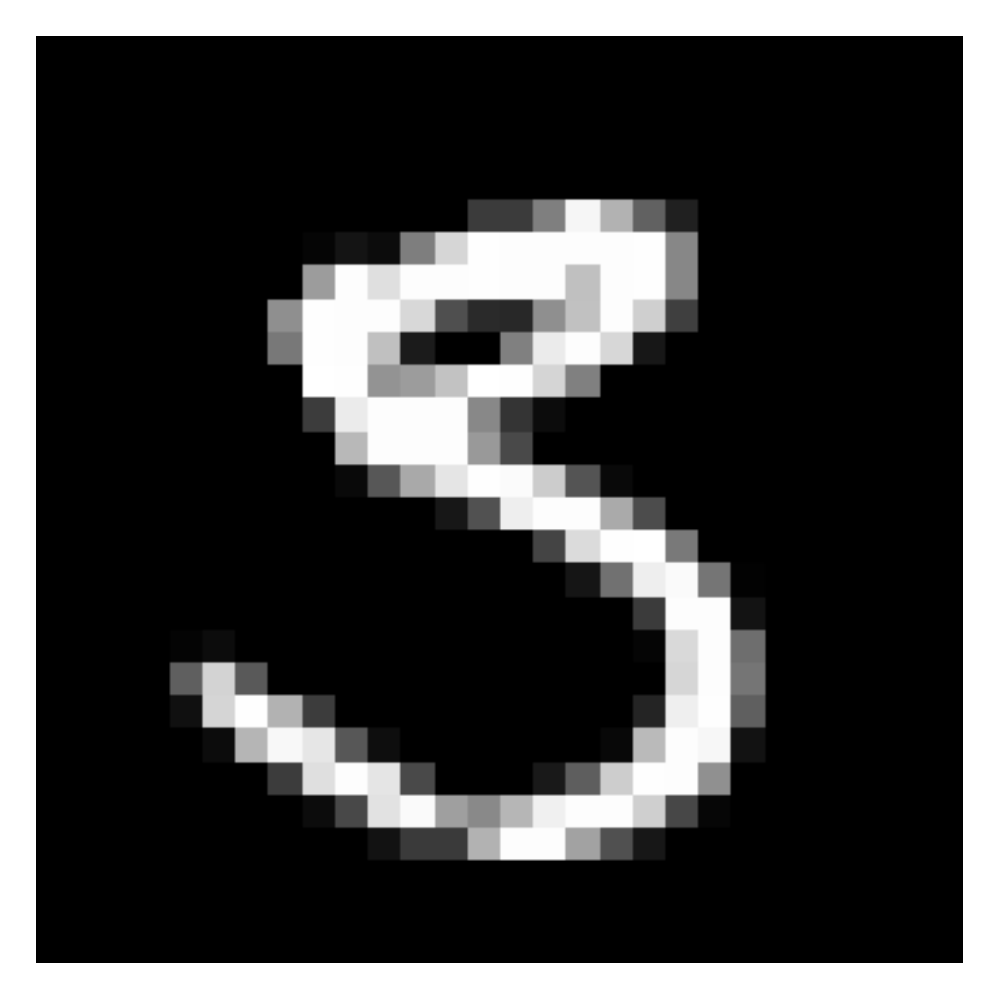}

\end{subfigure}
\begin{subfigure}[b]{0.105\linewidth}
\includegraphics[width=\linewidth]{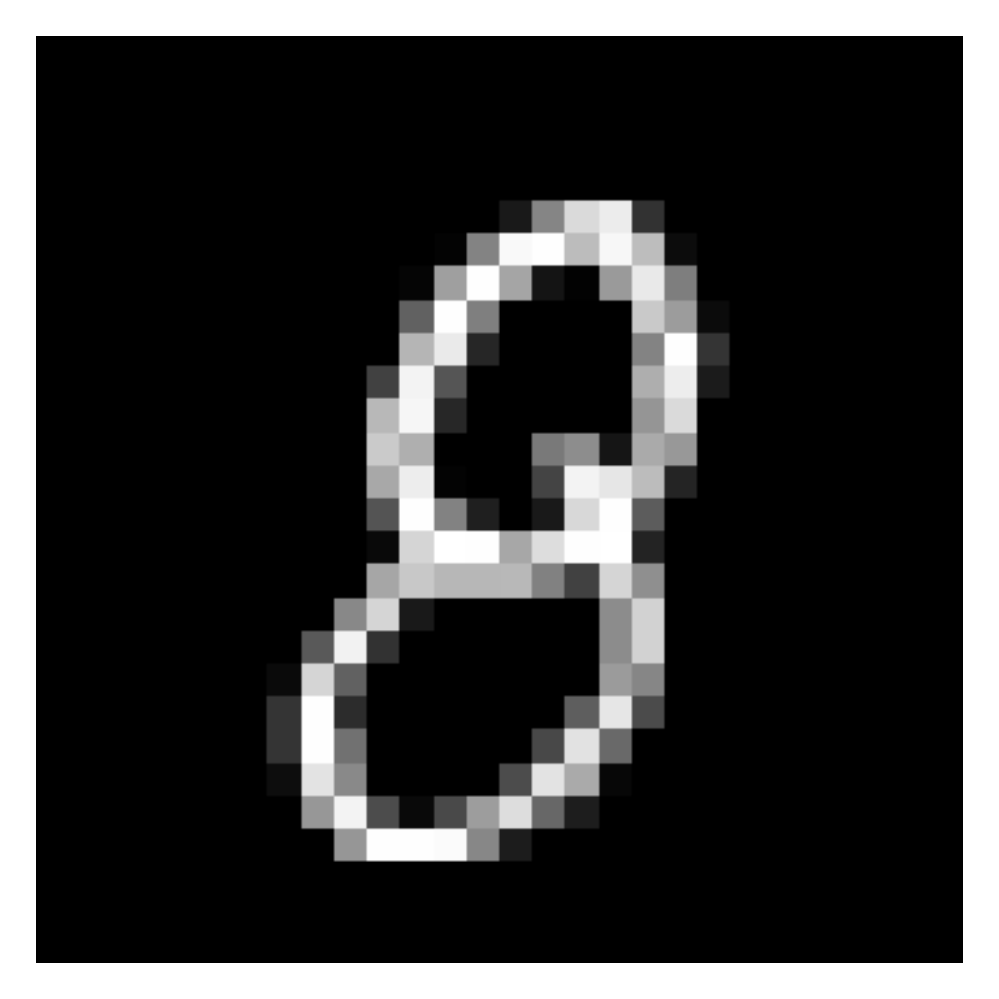}

\end{subfigure}

\begin{subfigure}[b]{0.105\linewidth}
\subcaption*{\textbf{3}: HC \newline  examples\newline }
\end{subfigure}
\begin{subfigure}[b]{0.105\linewidth}
\includegraphics[width=\linewidth]{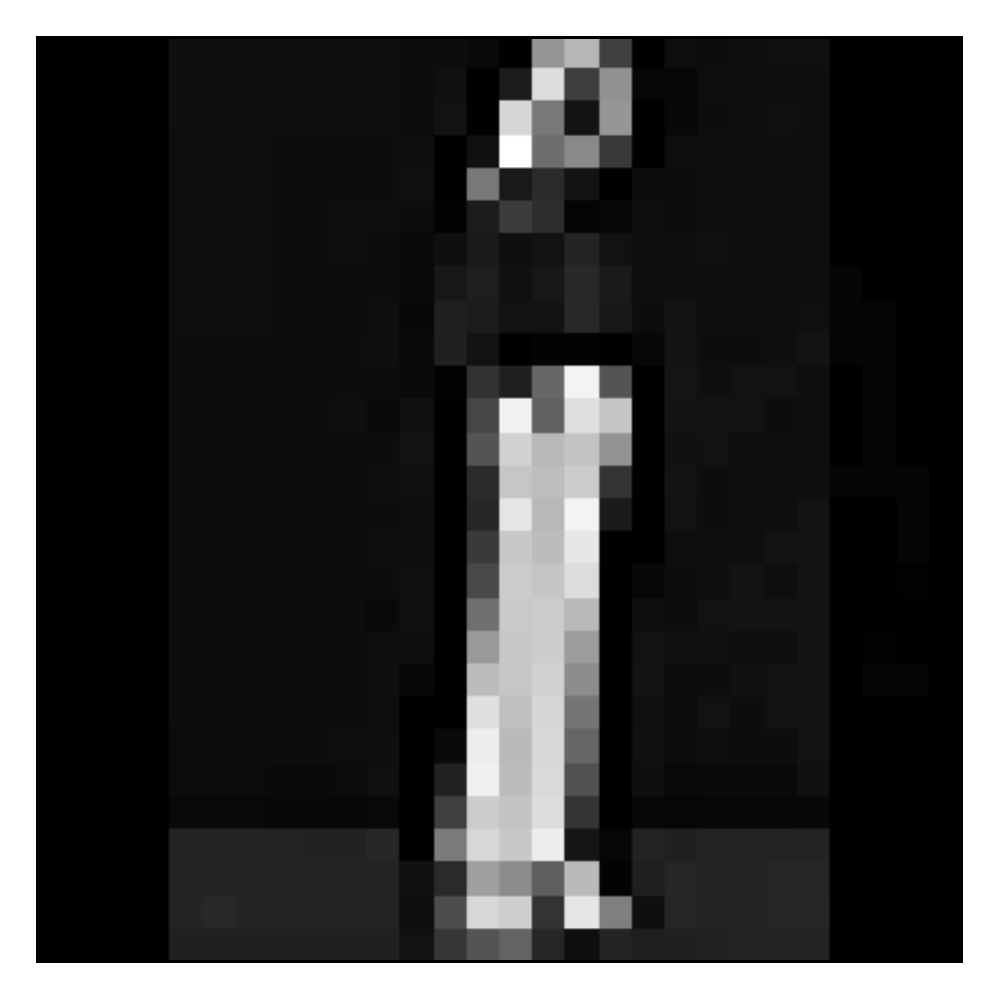}

\end{subfigure}
\begin{subfigure}[b]{0.105\linewidth}
\includegraphics[width=\linewidth]{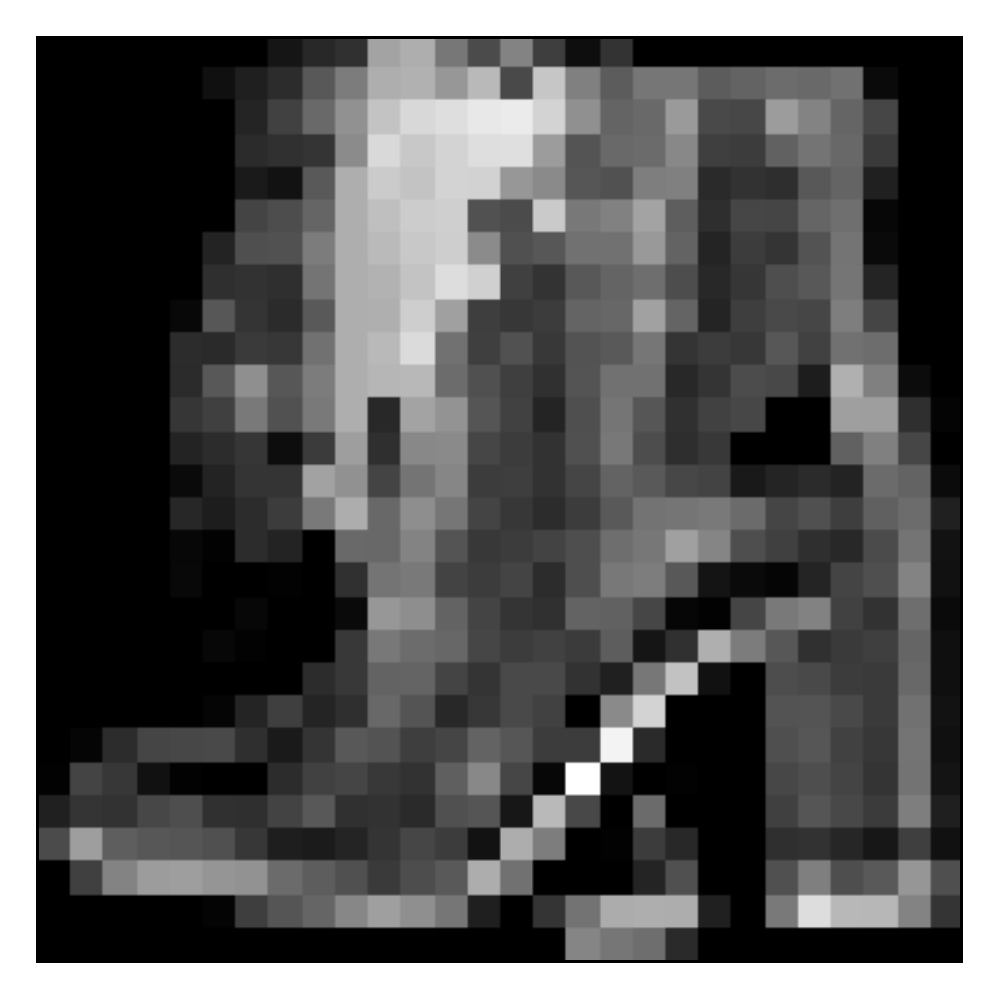}

\end{subfigure}
\begin{subfigure}[b]{0.105\linewidth}
\includegraphics[width=\linewidth]{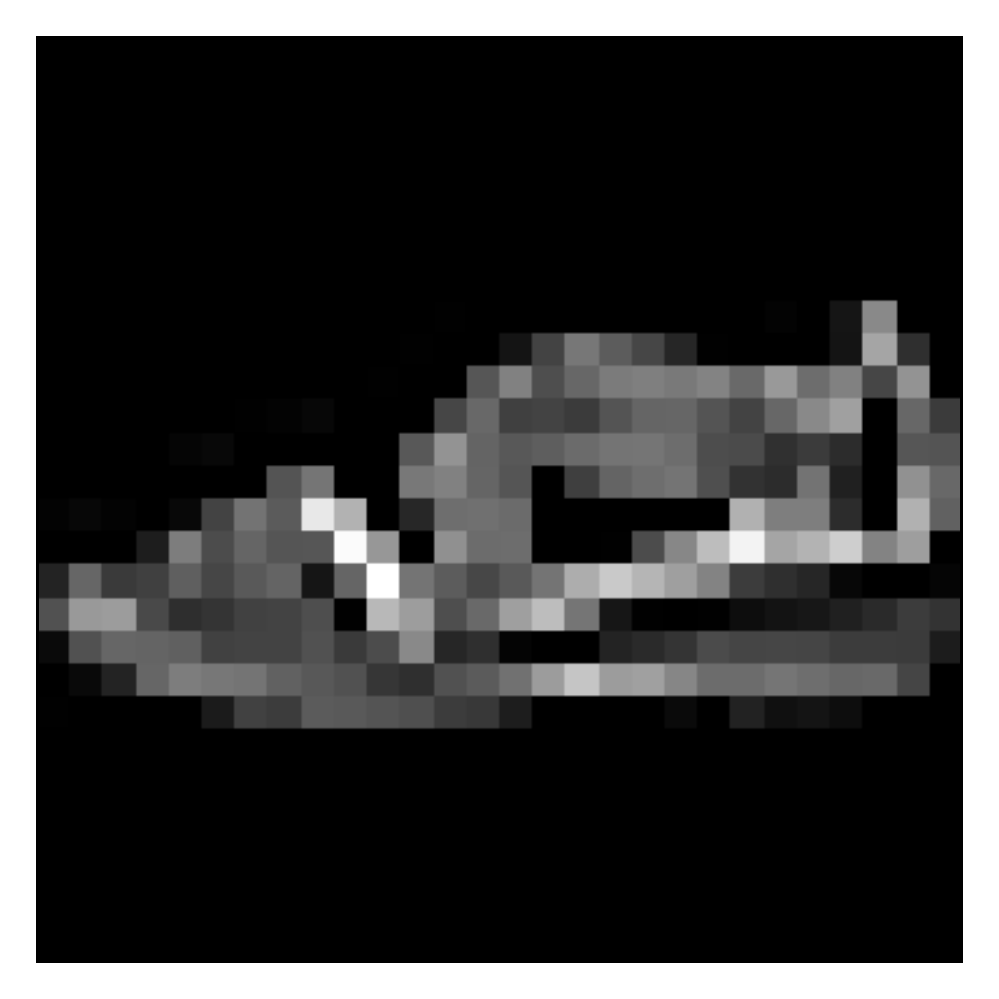}

\end{subfigure}
\begin{subfigure}[b]{0.105\linewidth}
\includegraphics[width=\linewidth]{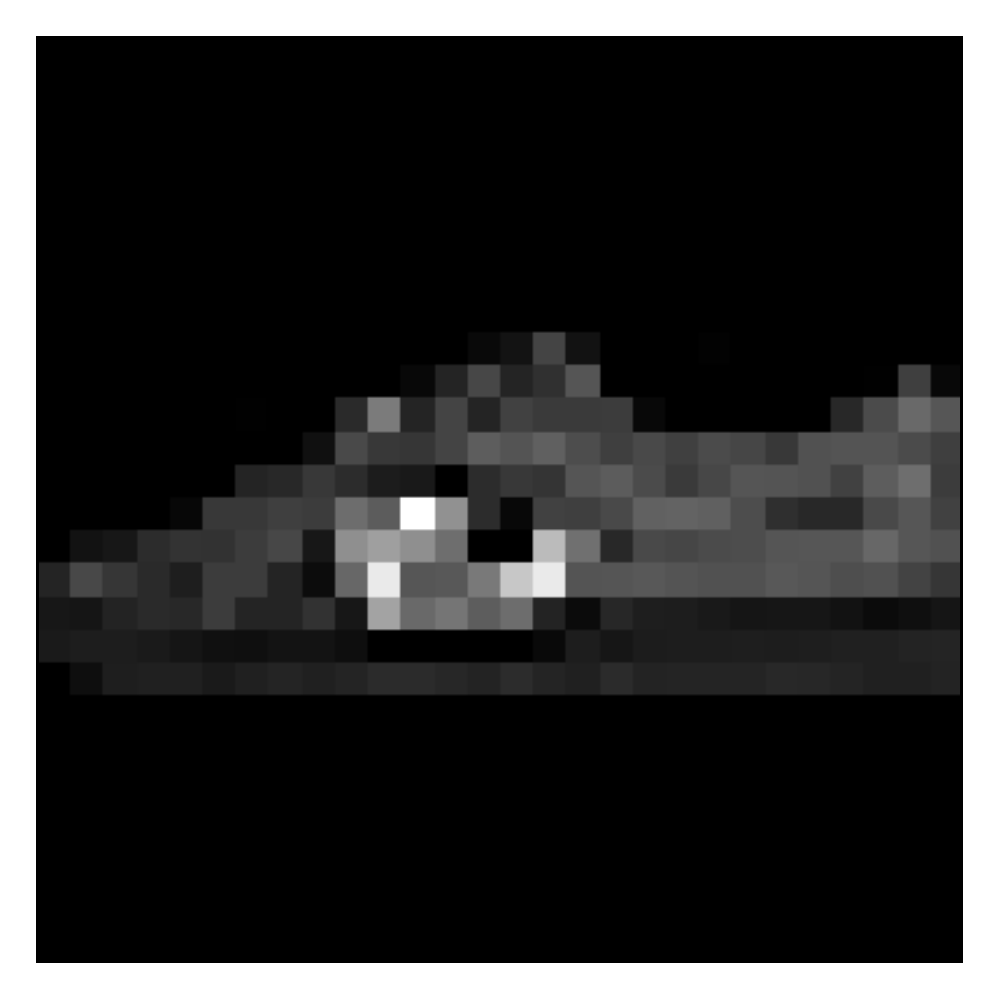}

\end{subfigure}
\begin{subfigure}[b]{0.105\linewidth}
\includegraphics[width=\linewidth]{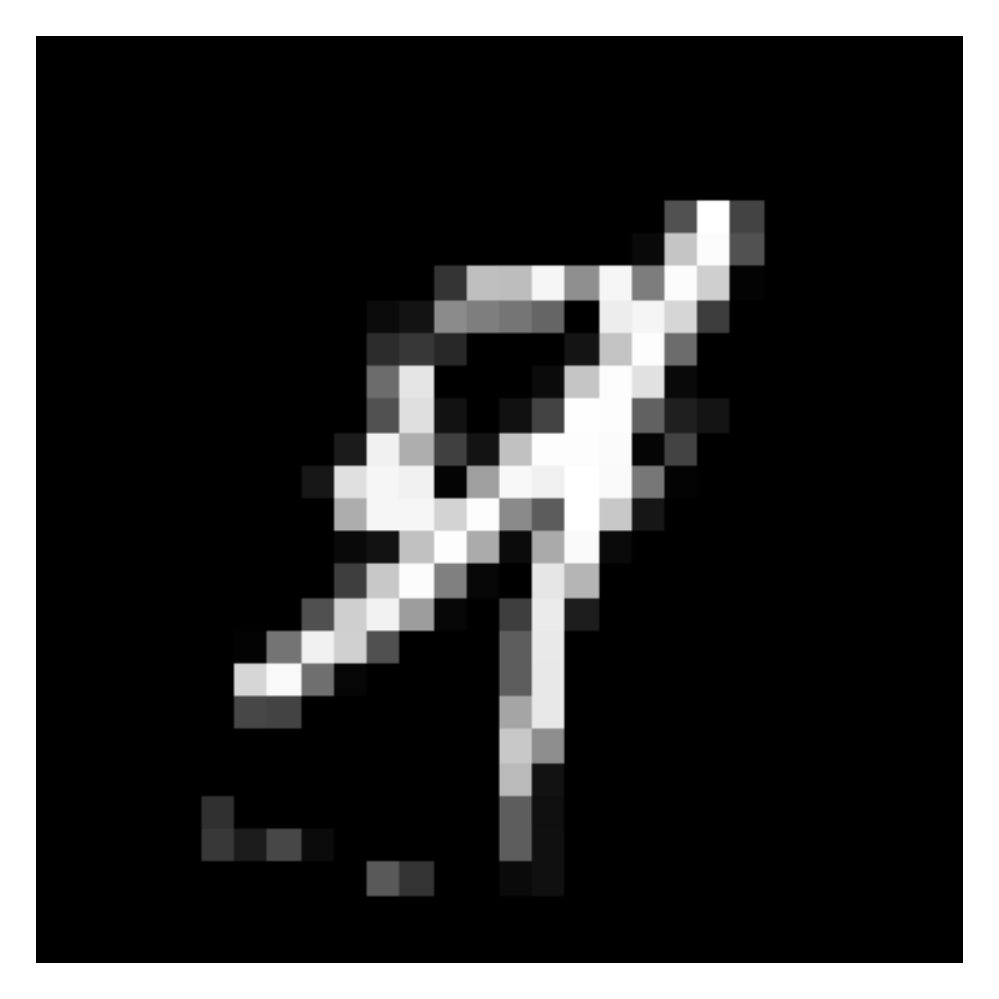}

\end{subfigure}
\begin{subfigure}[b]{0.105\linewidth}
\includegraphics[width=\linewidth]{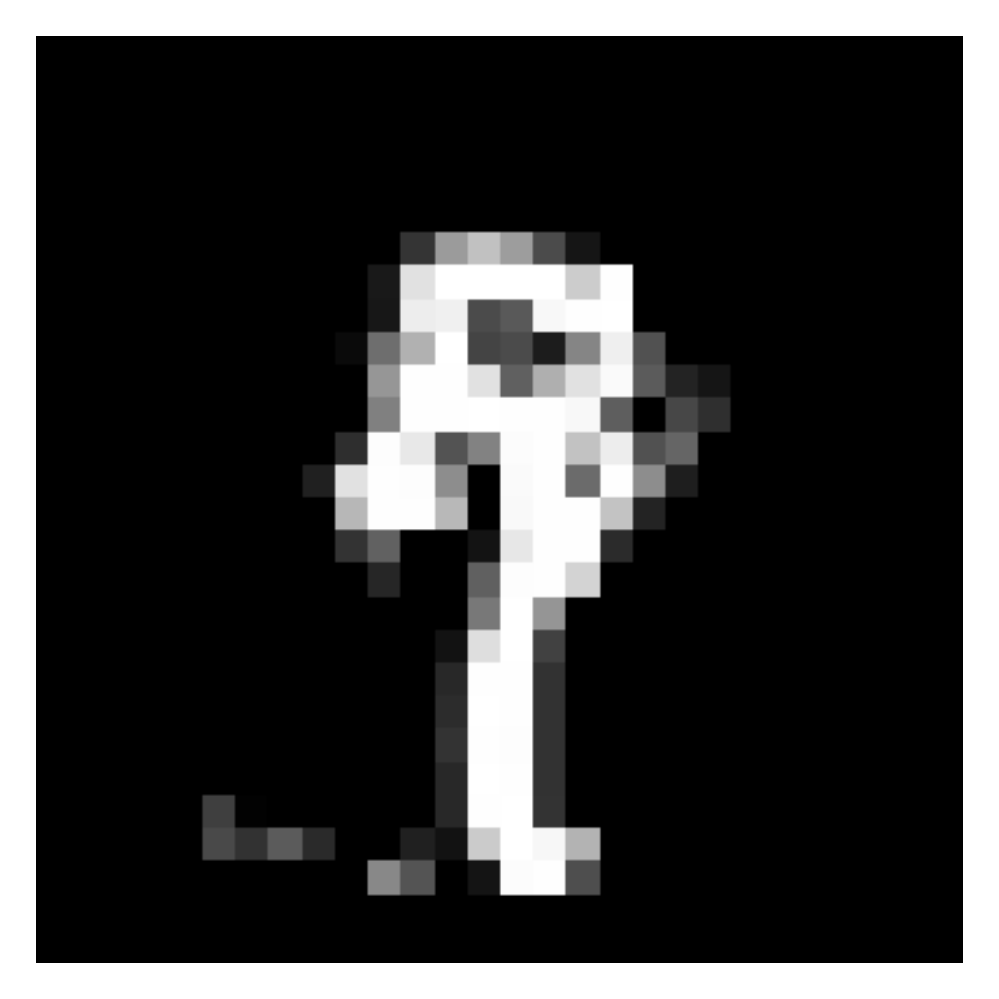}

\end{subfigure}
\begin{subfigure}[b]{0.105\linewidth}
\includegraphics[width=\linewidth]{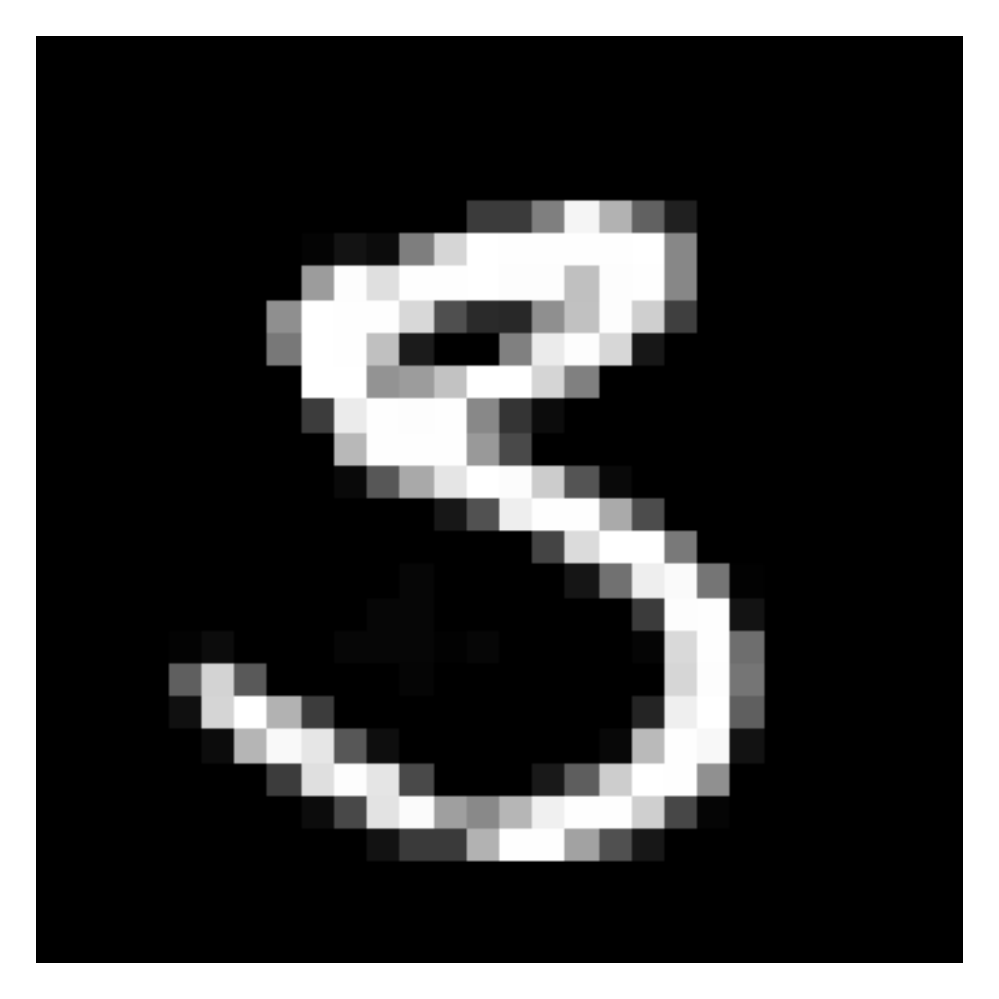}

\end{subfigure}
\begin{subfigure}[b]{0.105\linewidth}
\includegraphics[width=\linewidth]{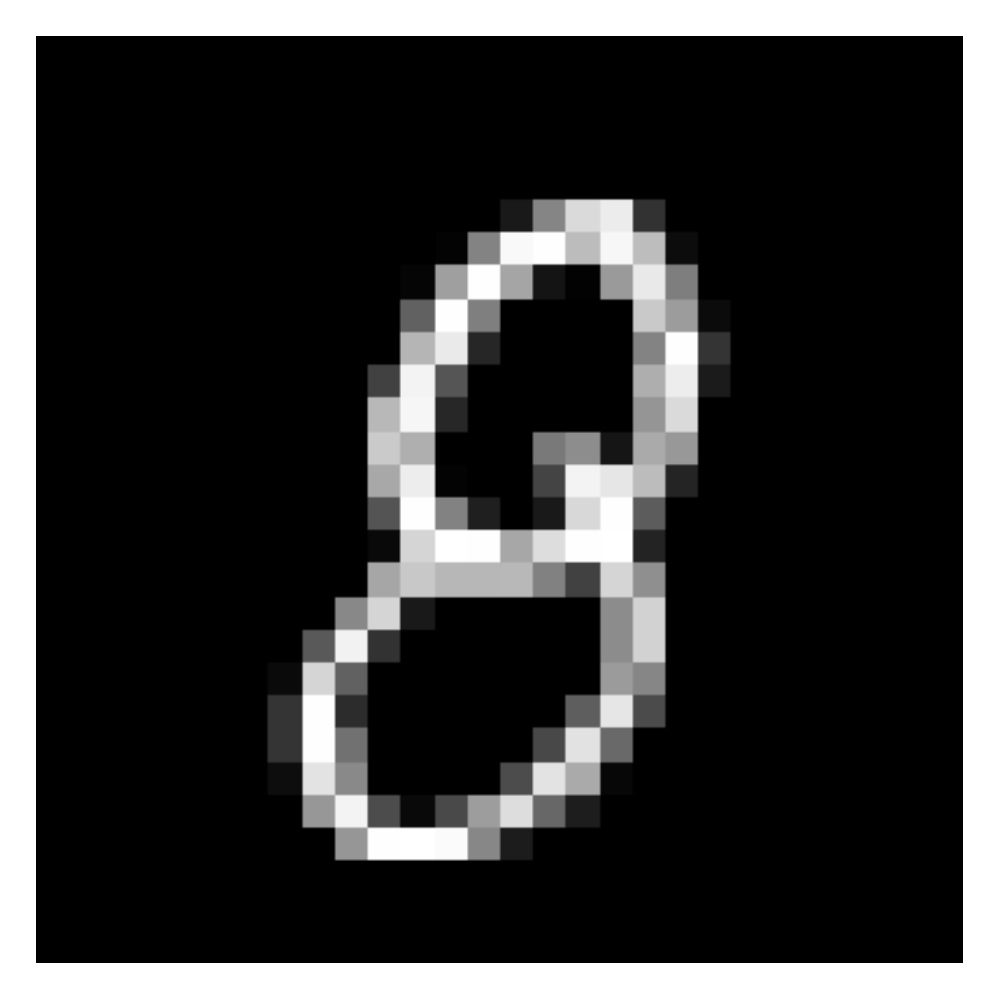}

\end{subfigure}

\begin{subfigure}[b]{0.105\linewidth}
\subcaption*{\textbf{4}: HC vs. \newline HCLU \newline }
\end{subfigure}
\begin{subfigure}[b]{0.105\linewidth}
\includegraphics[width=\linewidth]{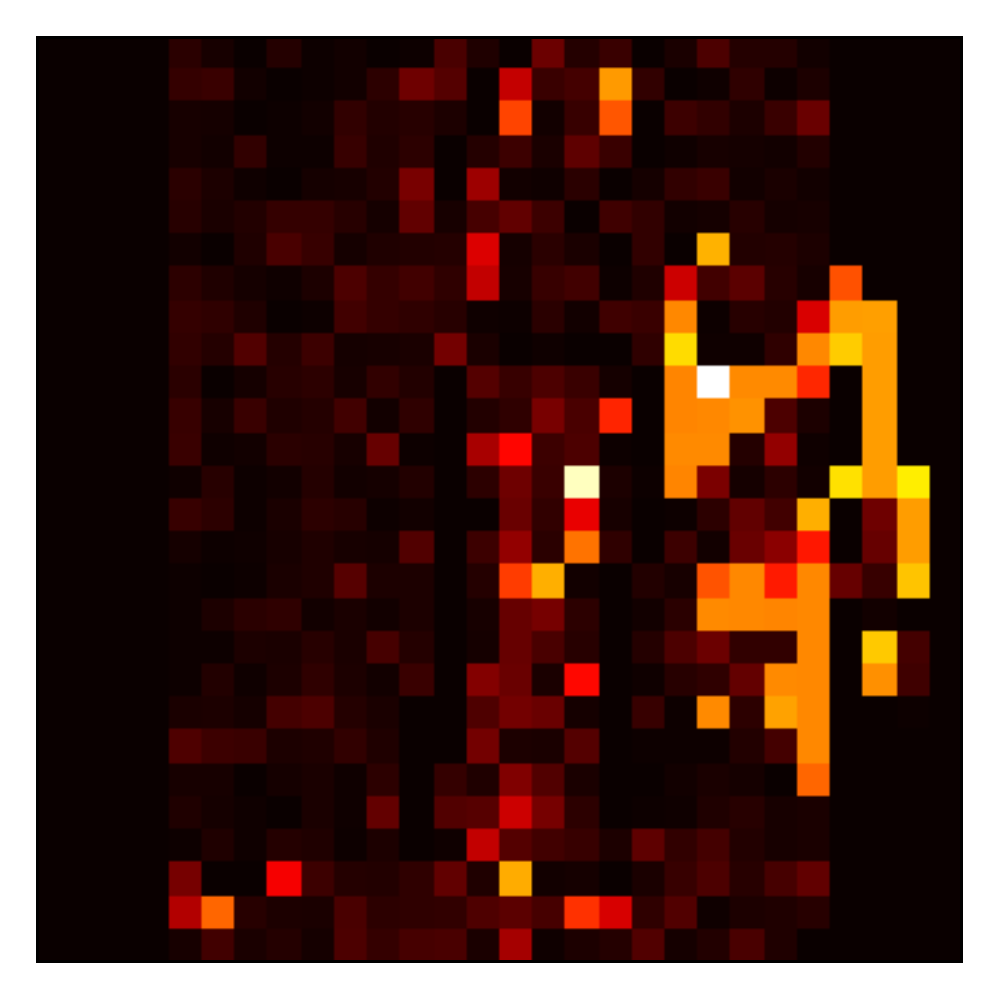}
\end{subfigure}
\begin{subfigure}[b]{0.105\linewidth}
\includegraphics[width=\linewidth]{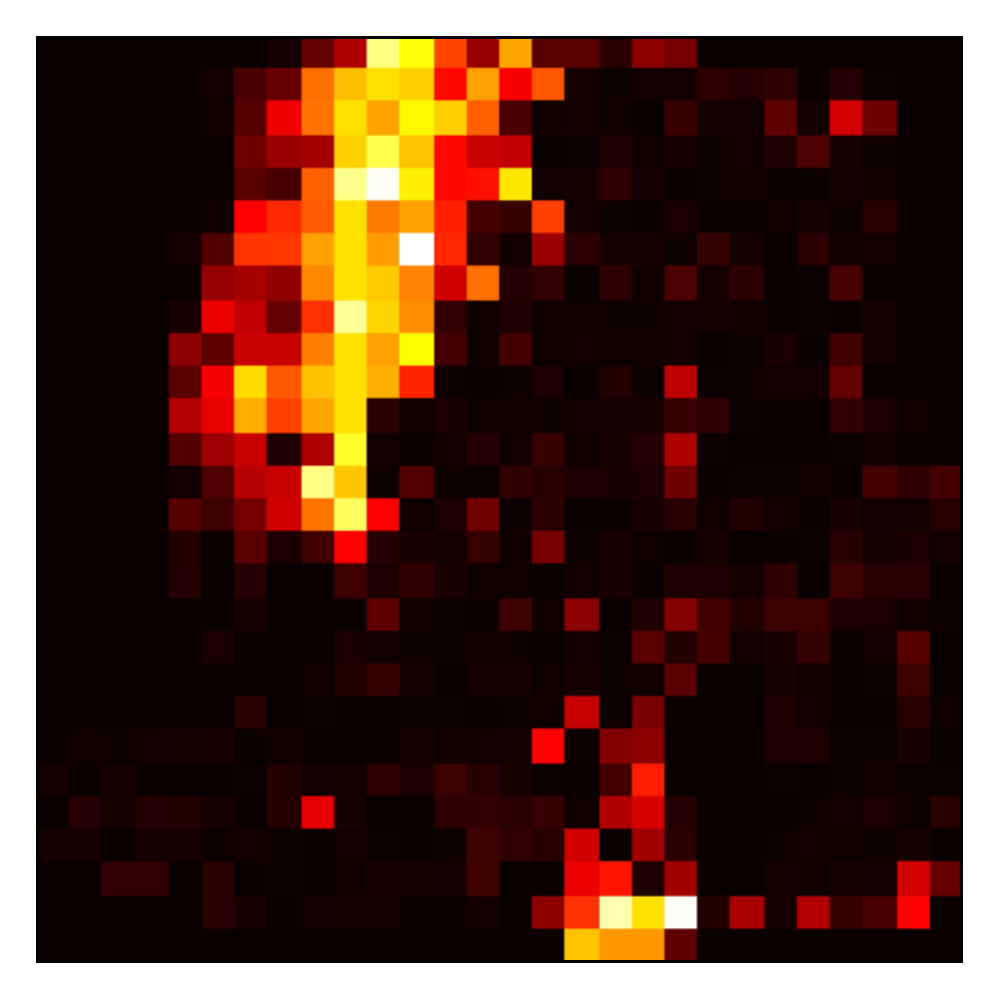}
\end{subfigure}
\begin{subfigure}[b]{0.105\linewidth}
\includegraphics[width=\linewidth]{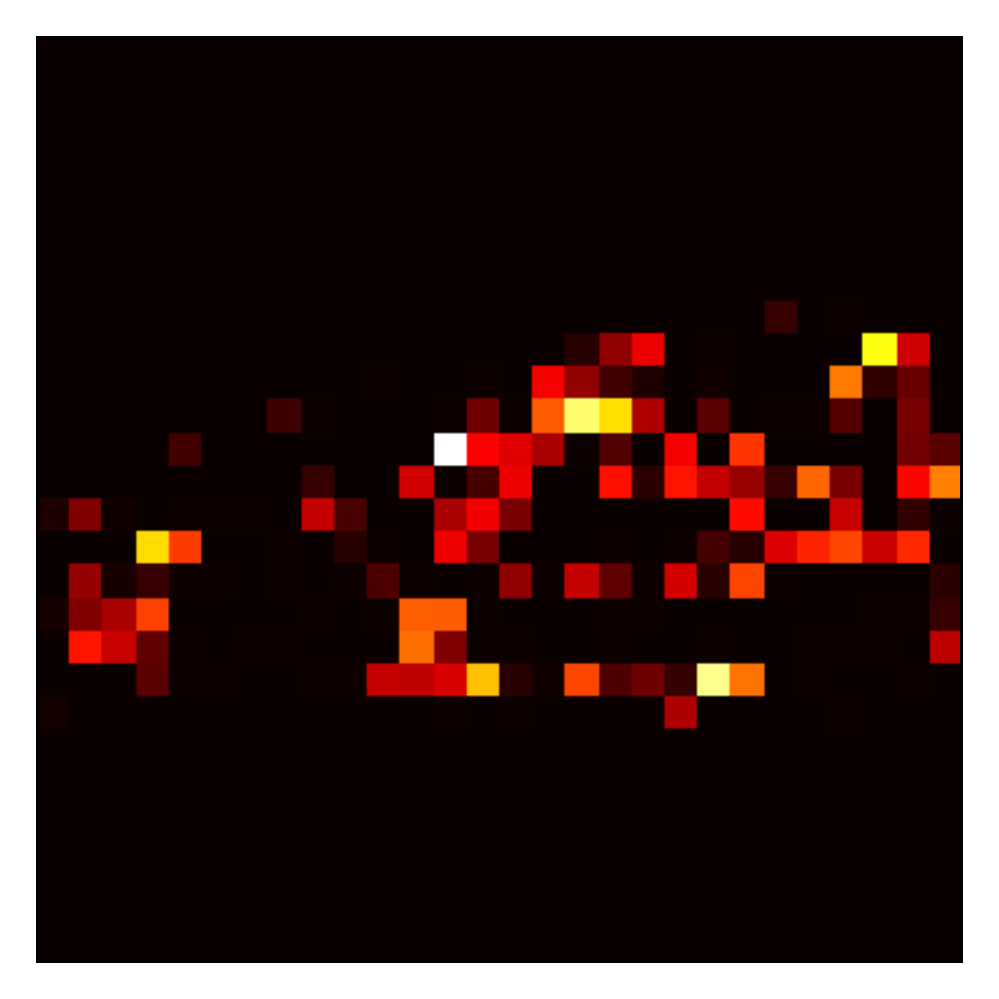}
\end{subfigure}
\begin{subfigure}[b]{0.105\linewidth}
\includegraphics[width=\linewidth]{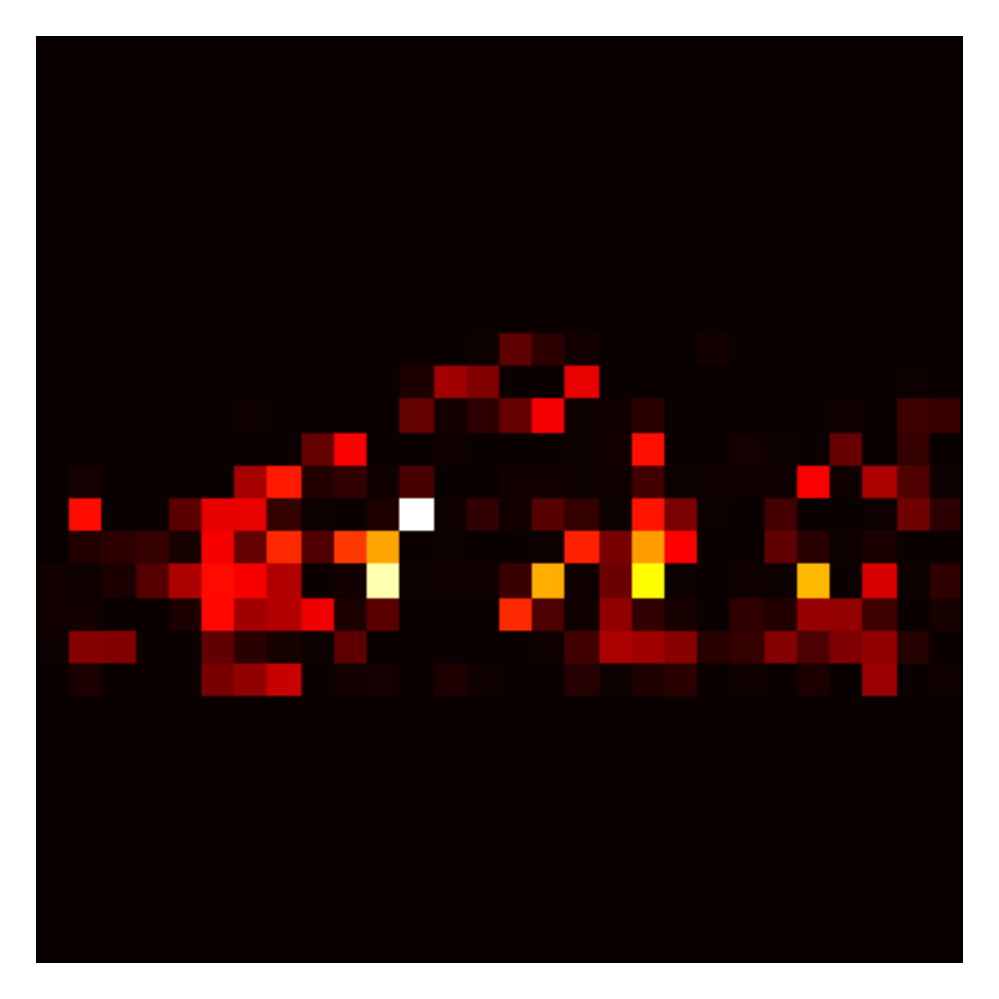}
\end{subfigure}
\begin{subfigure}[b]{0.105\linewidth}
\includegraphics[width=\linewidth]{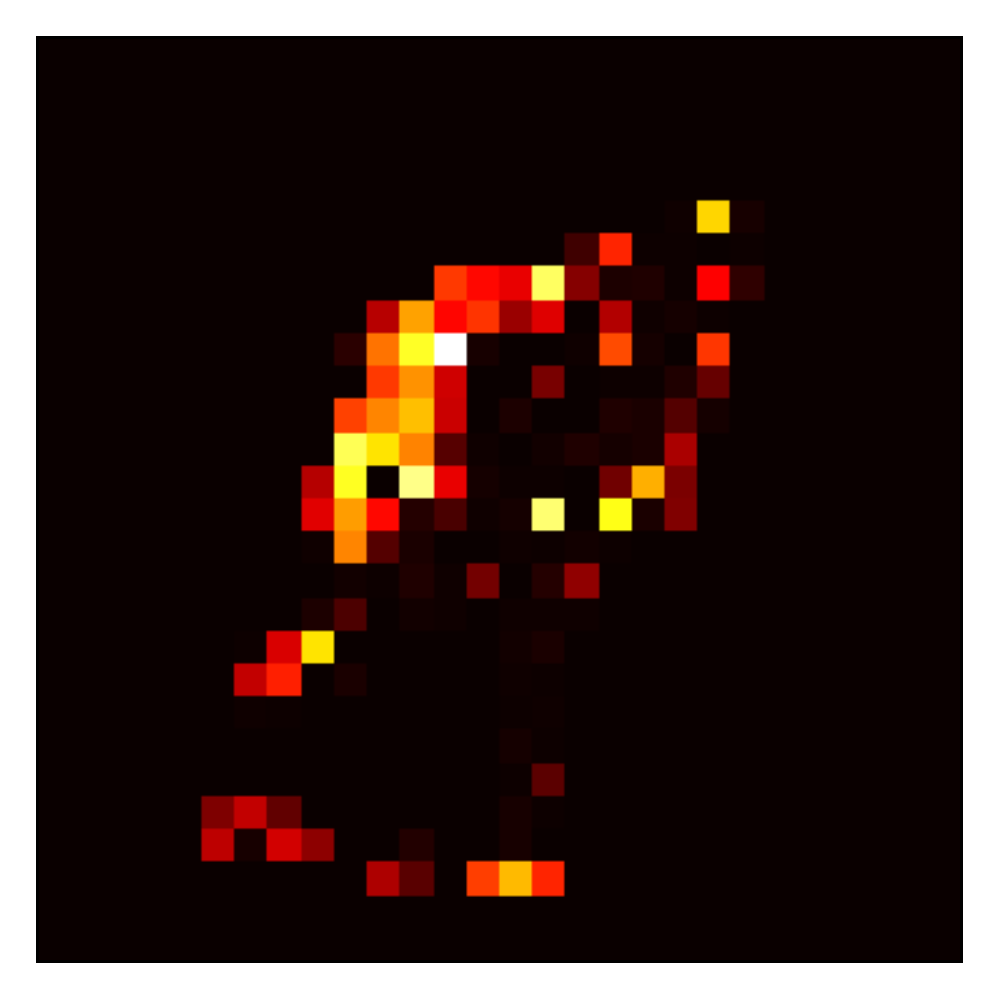}
\end{subfigure}
\begin{subfigure}[b]{0.105\linewidth}
\includegraphics[width=\linewidth]{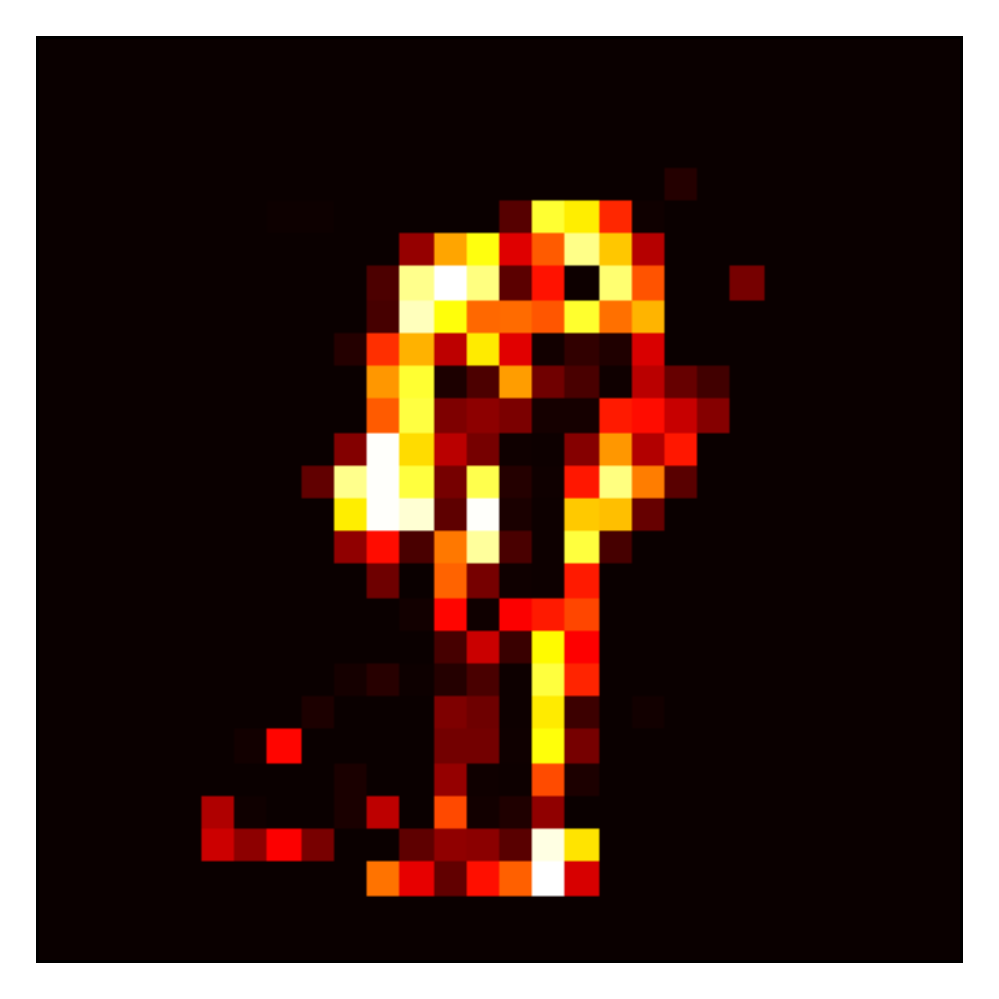}
\end{subfigure}
\begin{subfigure}[b]{0.105\linewidth}
\includegraphics[width=\linewidth]{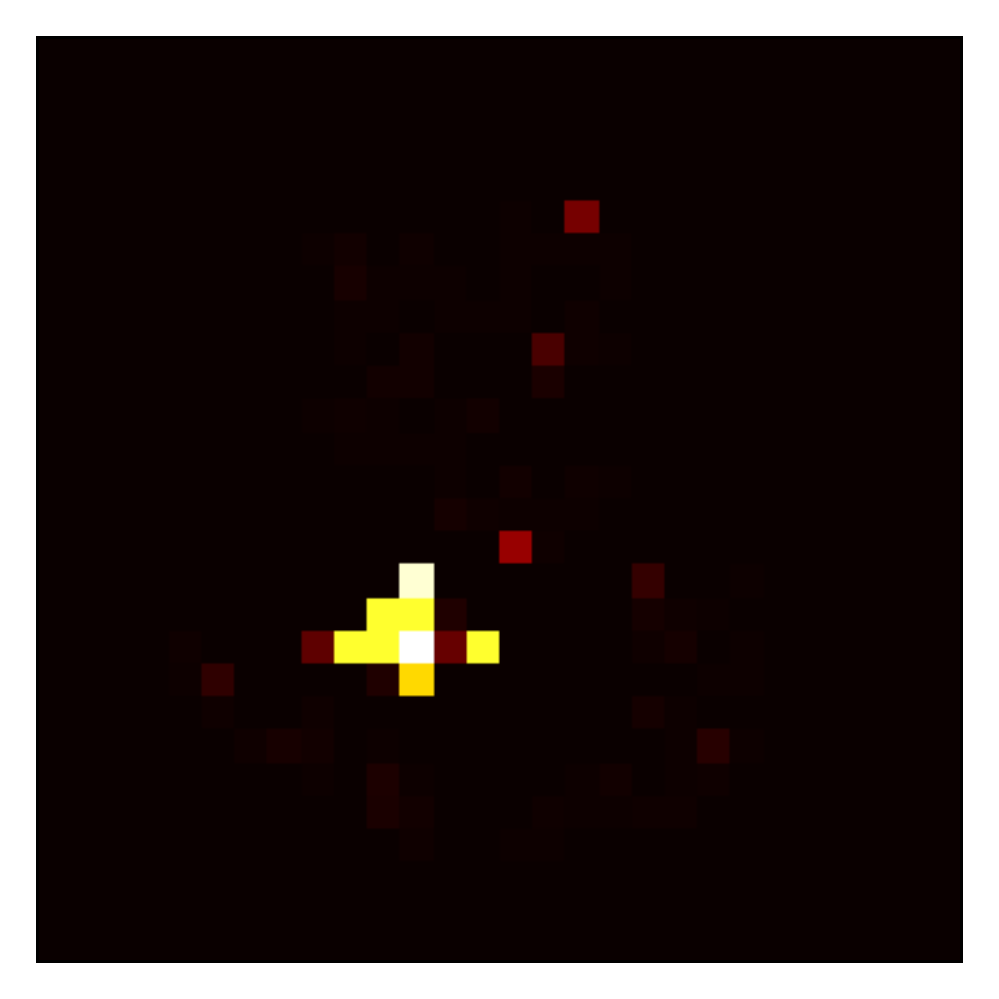}
\end{subfigure}
\begin{subfigure}[b]{0.105\linewidth}
\includegraphics[width=\linewidth]{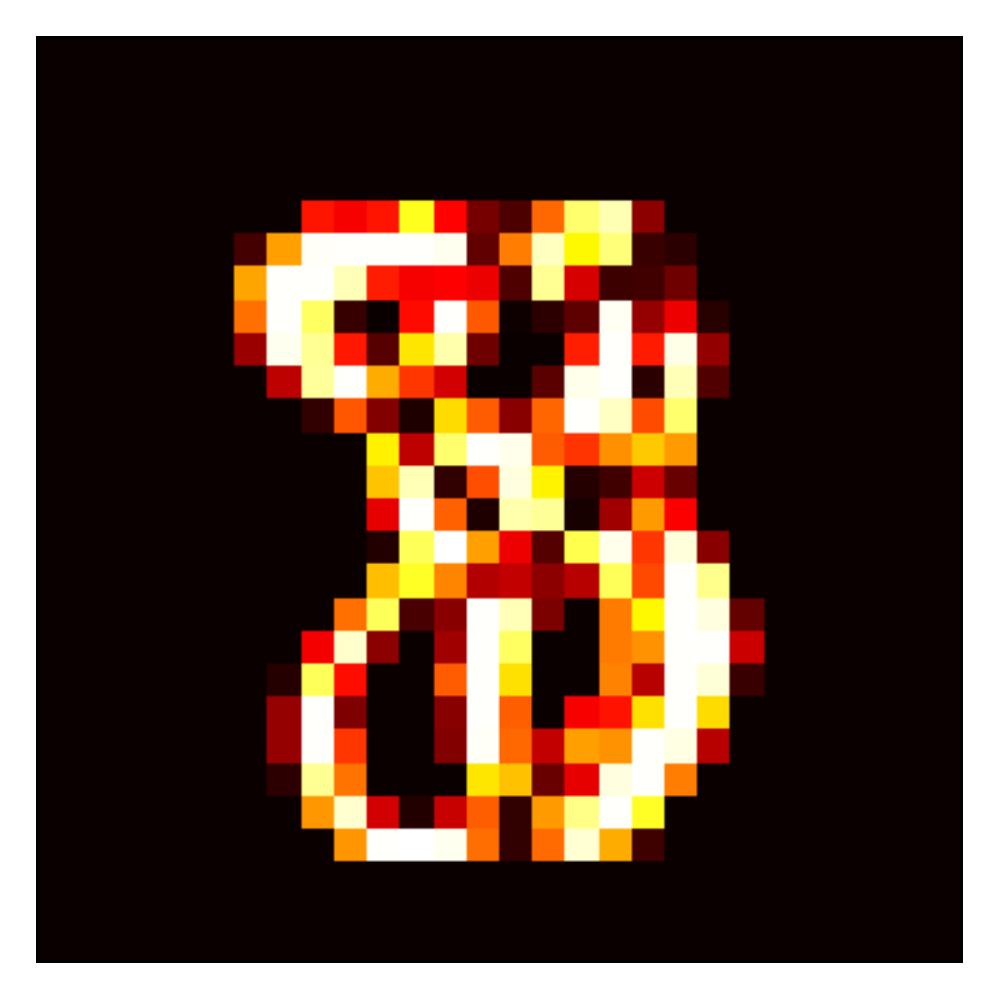}
\end{subfigure}

\begin{subfigure}[b]{0.105\linewidth}
\subcaption*{\textbf{5}: HCLU  \newline  examples \newline  }
\end{subfigure}
\begin{subfigure}[b]{0.105\linewidth}
\includegraphics[width=\linewidth]{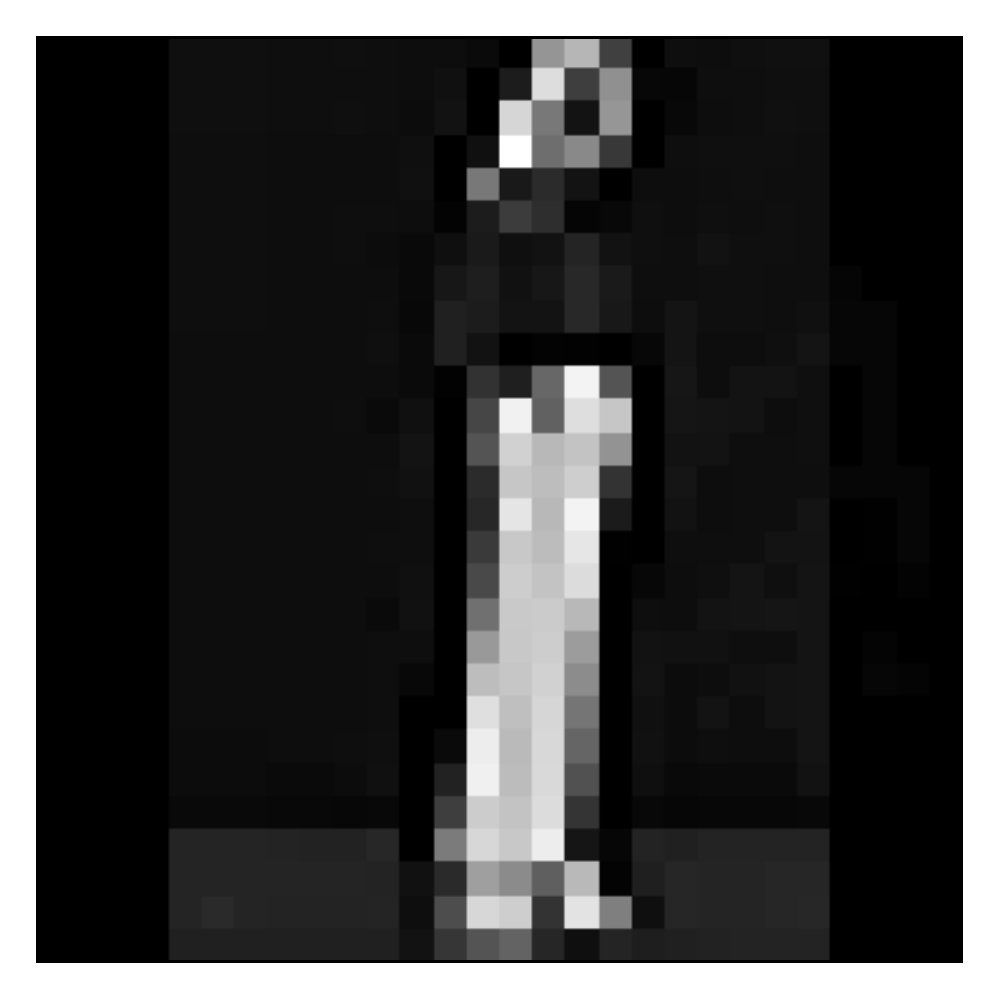}

\end{subfigure}
\begin{subfigure}[b]{0.105\linewidth}
\includegraphics[width=\linewidth]{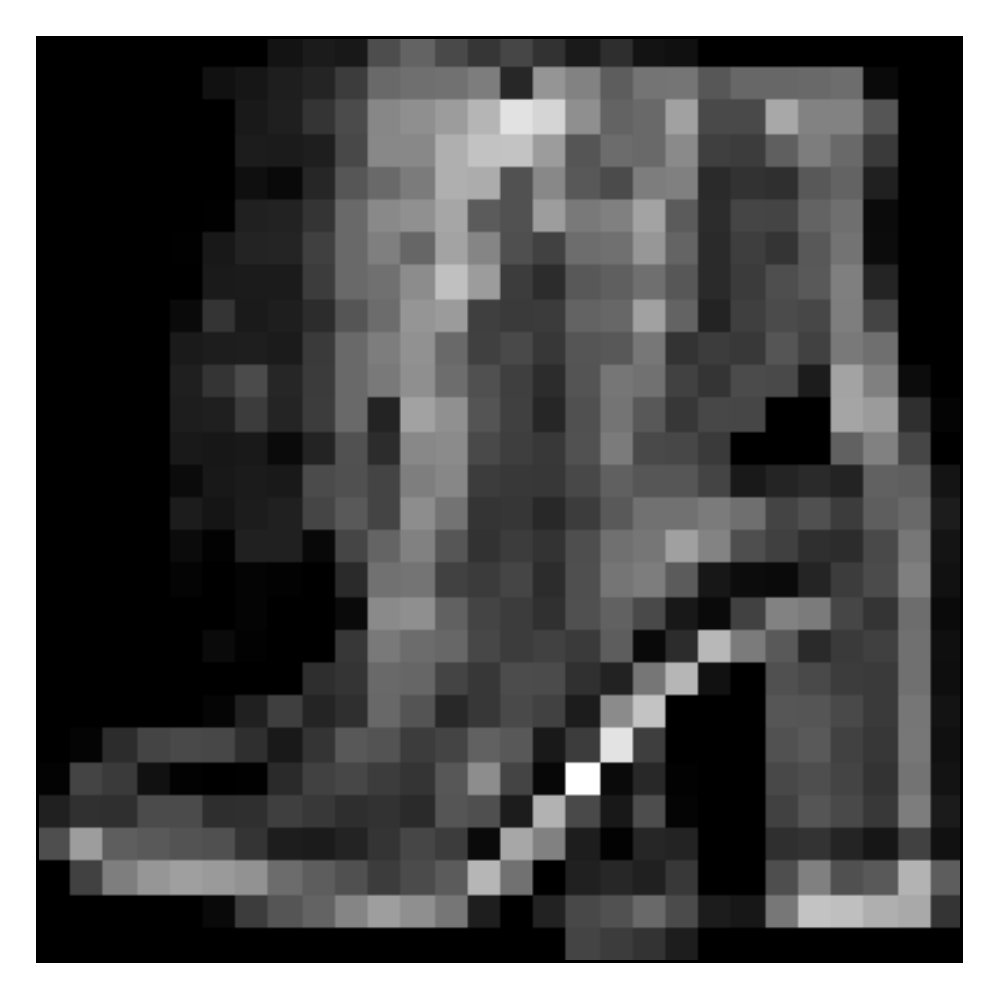}

\end{subfigure}
\begin{subfigure}[b]{0.105\linewidth}
\includegraphics[width=\linewidth]{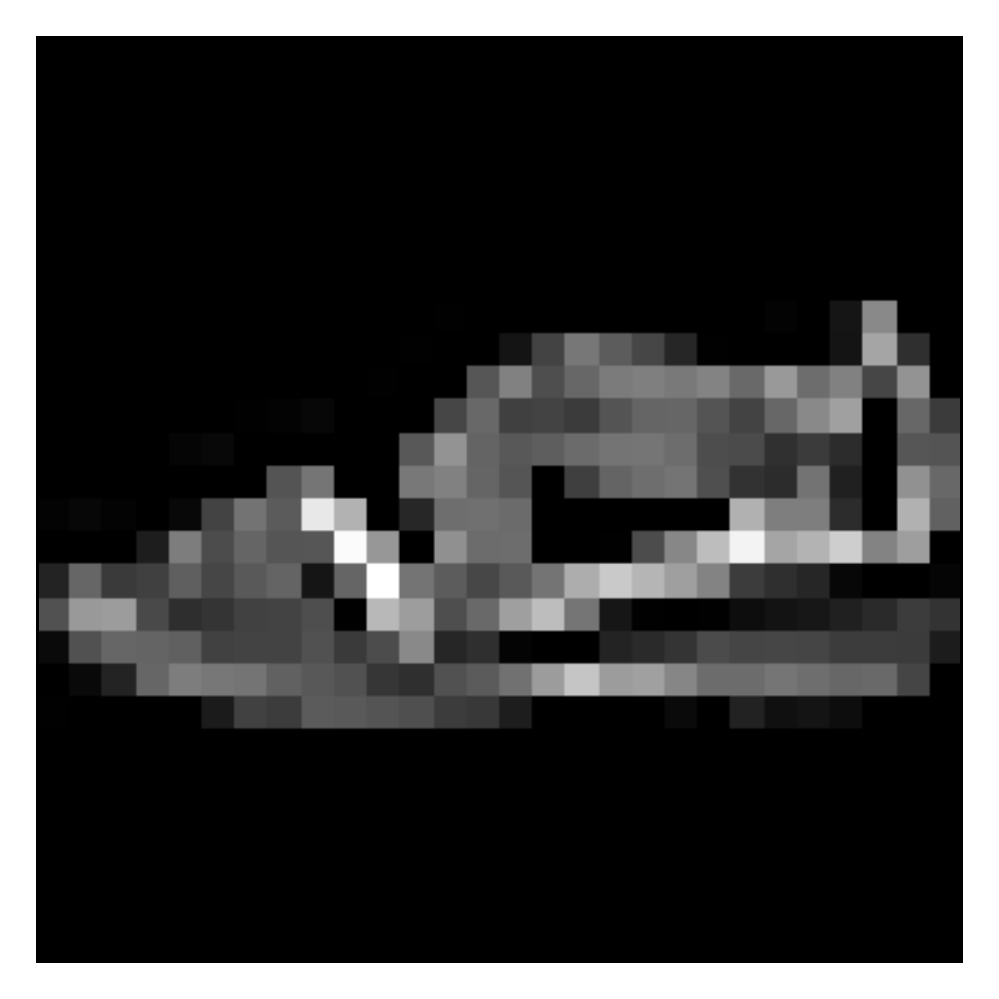}

\end{subfigure}
\begin{subfigure}[b]{0.105\linewidth}
\includegraphics[width=\linewidth]{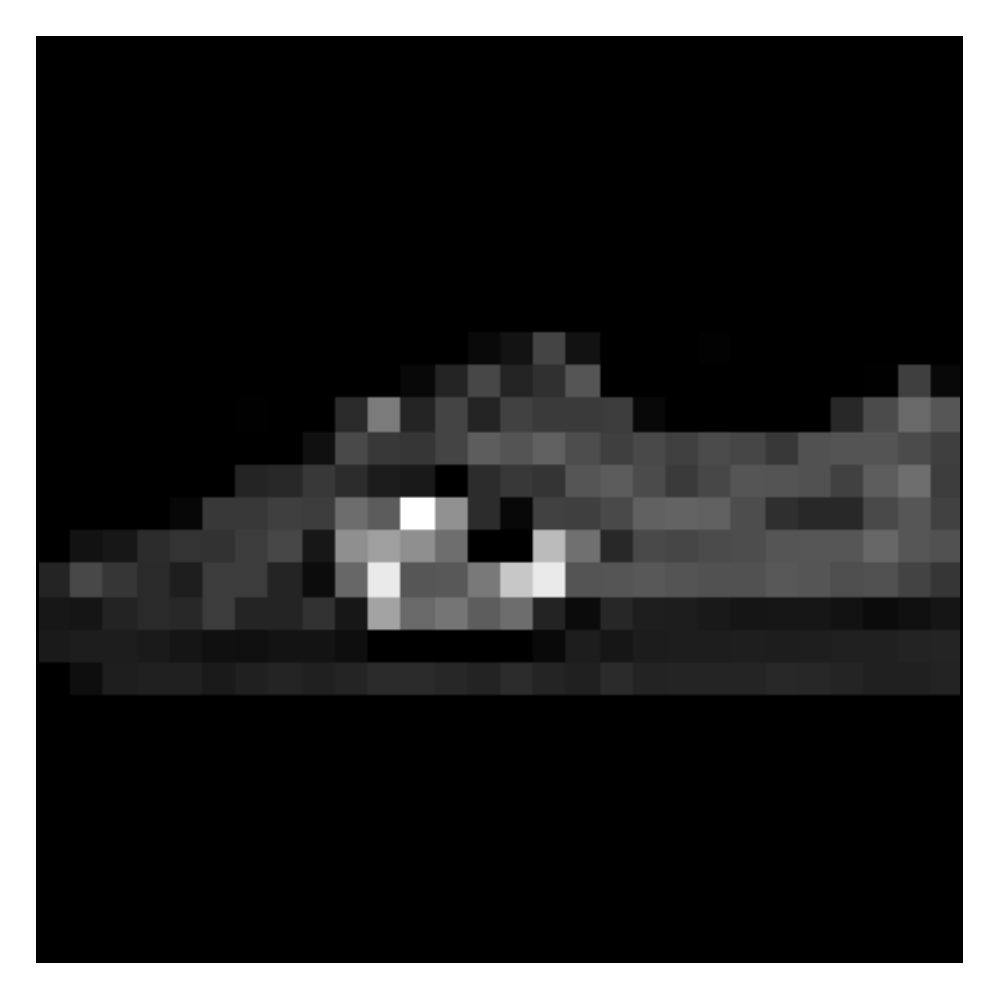}

\end{subfigure}
\begin{subfigure}[b]{0.105\linewidth}
\includegraphics[width=\linewidth]{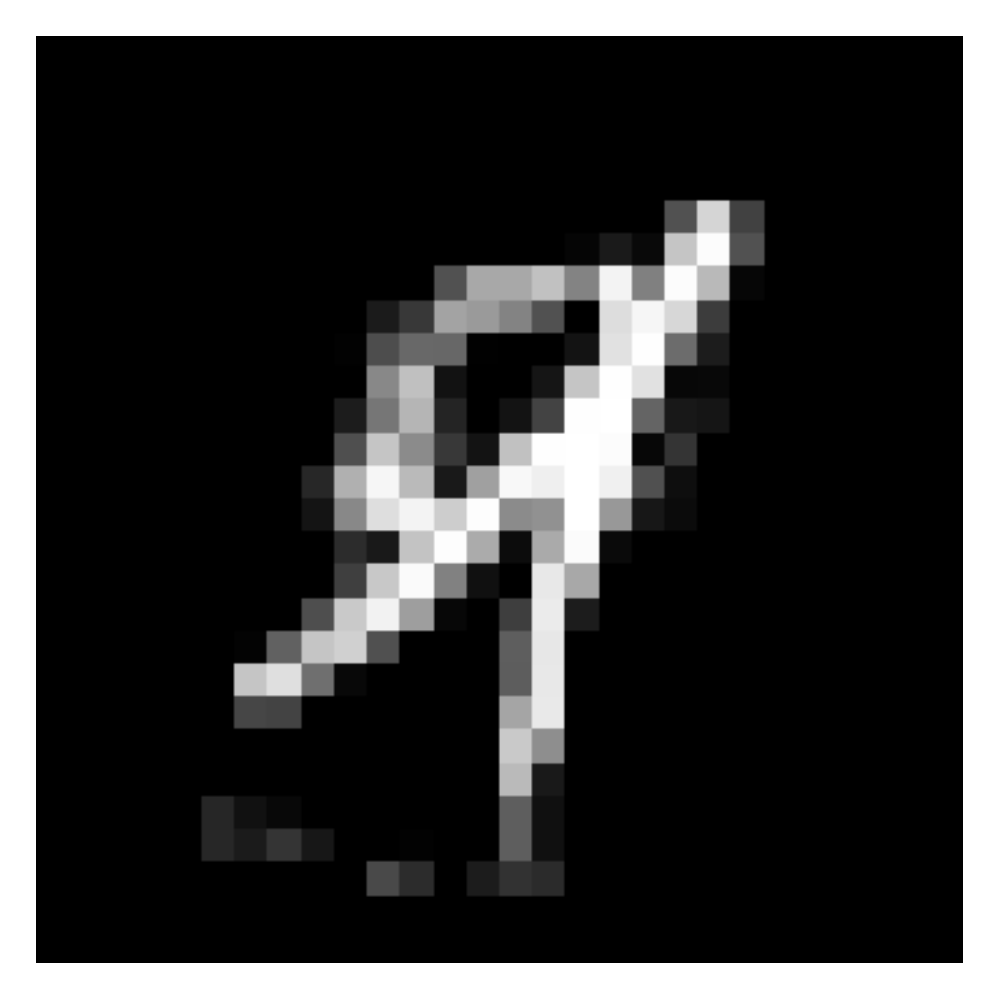}

\end{subfigure}
\begin{subfigure}[b]{0.105\linewidth}
\includegraphics[width=\linewidth]{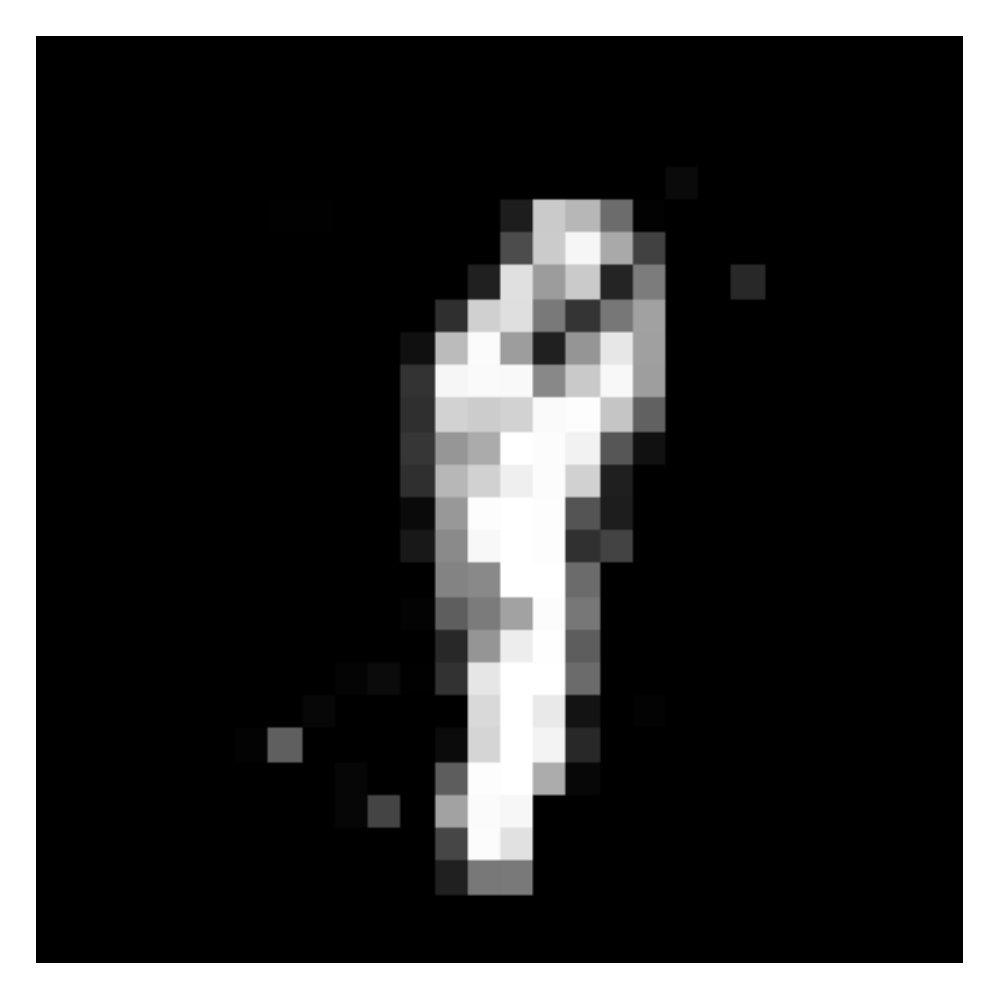}

\end{subfigure}
\begin{subfigure}[b]{0.105\linewidth}
\includegraphics[width=\linewidth]{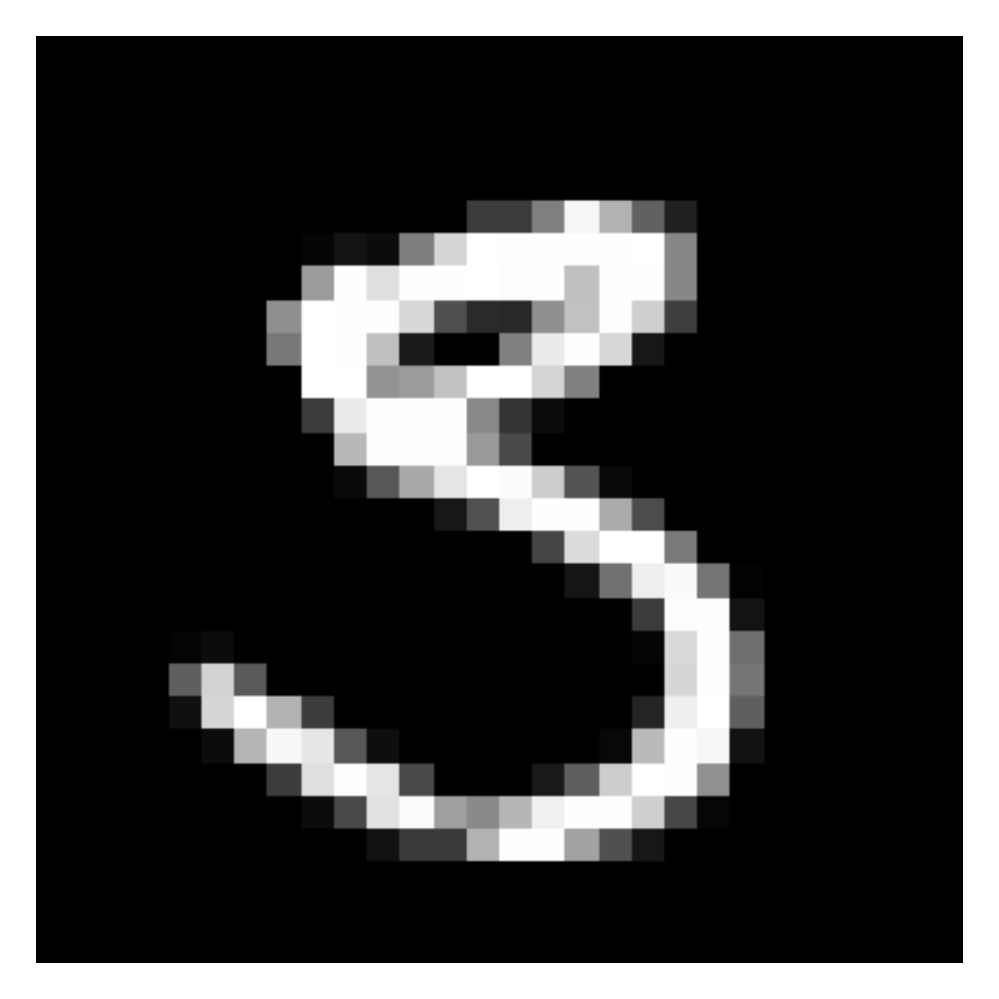}

\end{subfigure}
\begin{subfigure}[b]{0.105\linewidth}
\includegraphics[width=\linewidth]{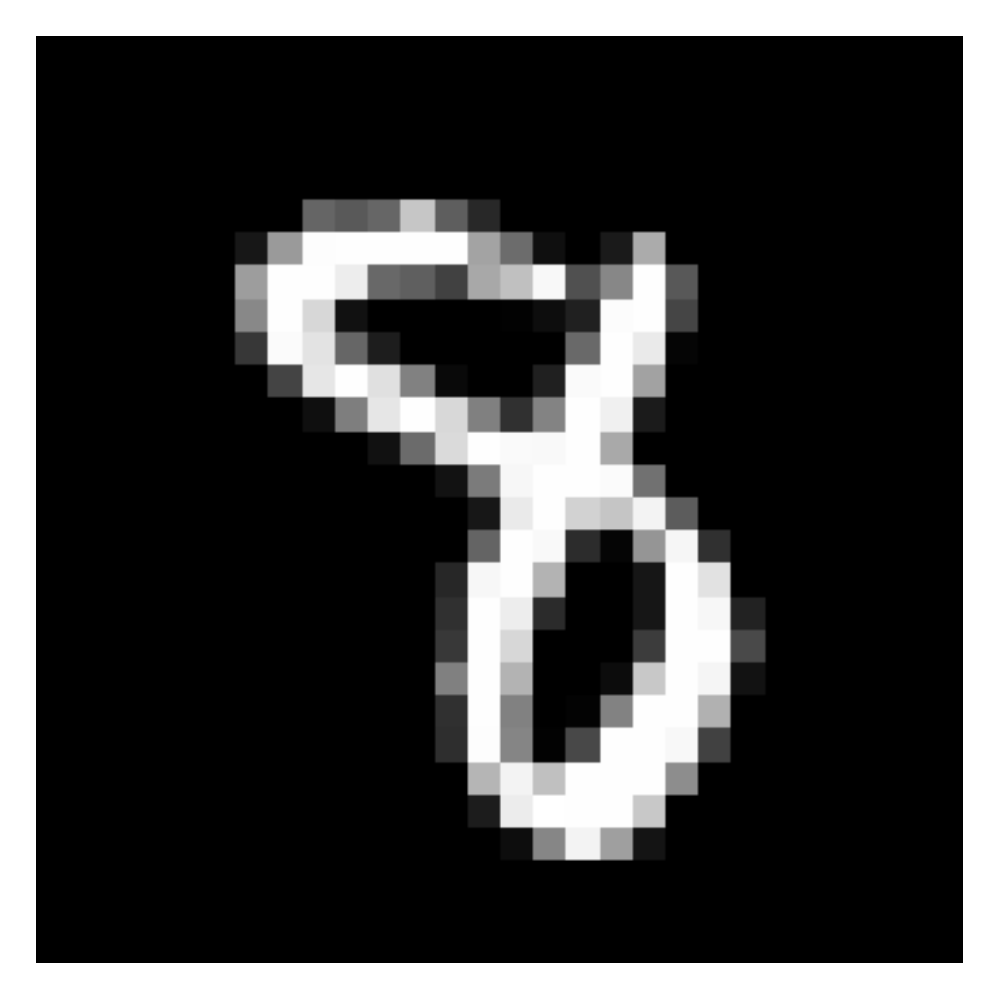}

\end{subfigure}

\begin{subfigure}[b]{0.105\linewidth}
\subcaption*{\textbf{6}: Benign \newline original \newline }
\end{subfigure}
\begin{subfigure}[b]{0.105\linewidth}
\includegraphics[width=\linewidth]{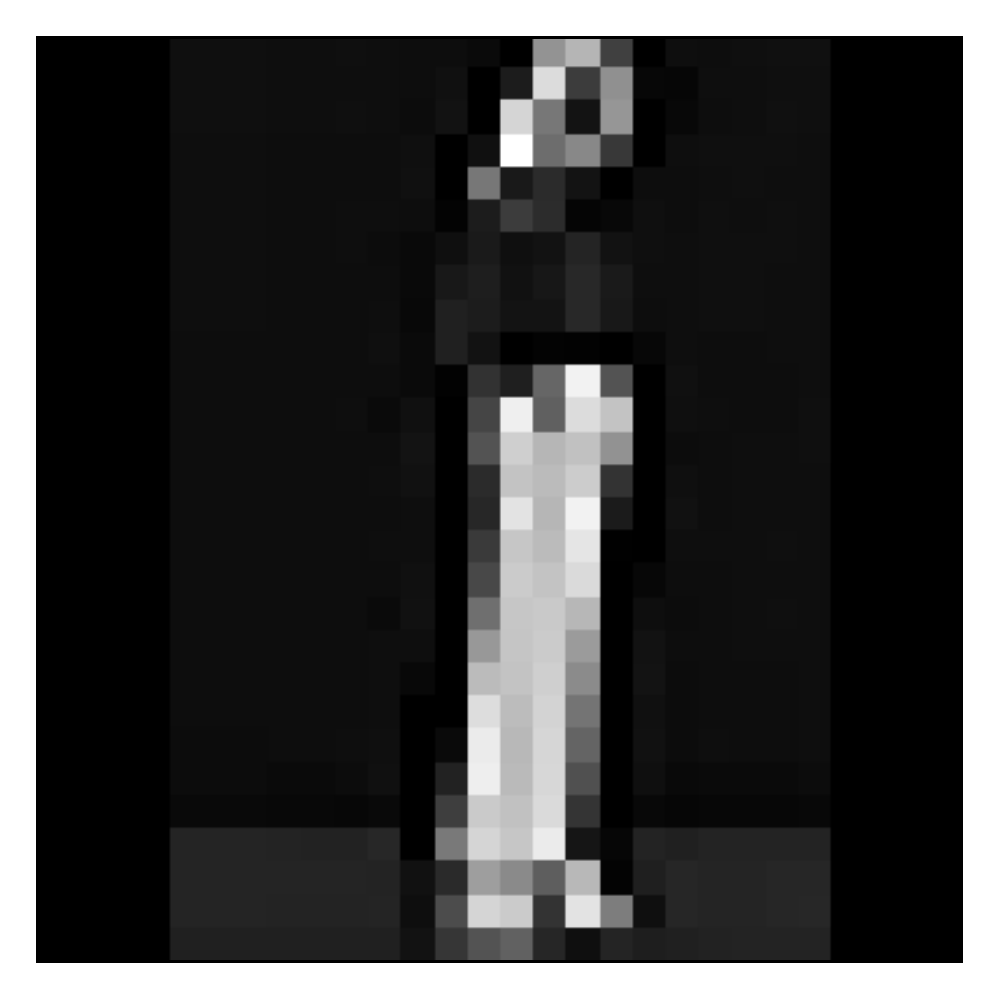}

\end{subfigure}
\begin{subfigure}[b]{0.105\linewidth}
\includegraphics[width=\linewidth]{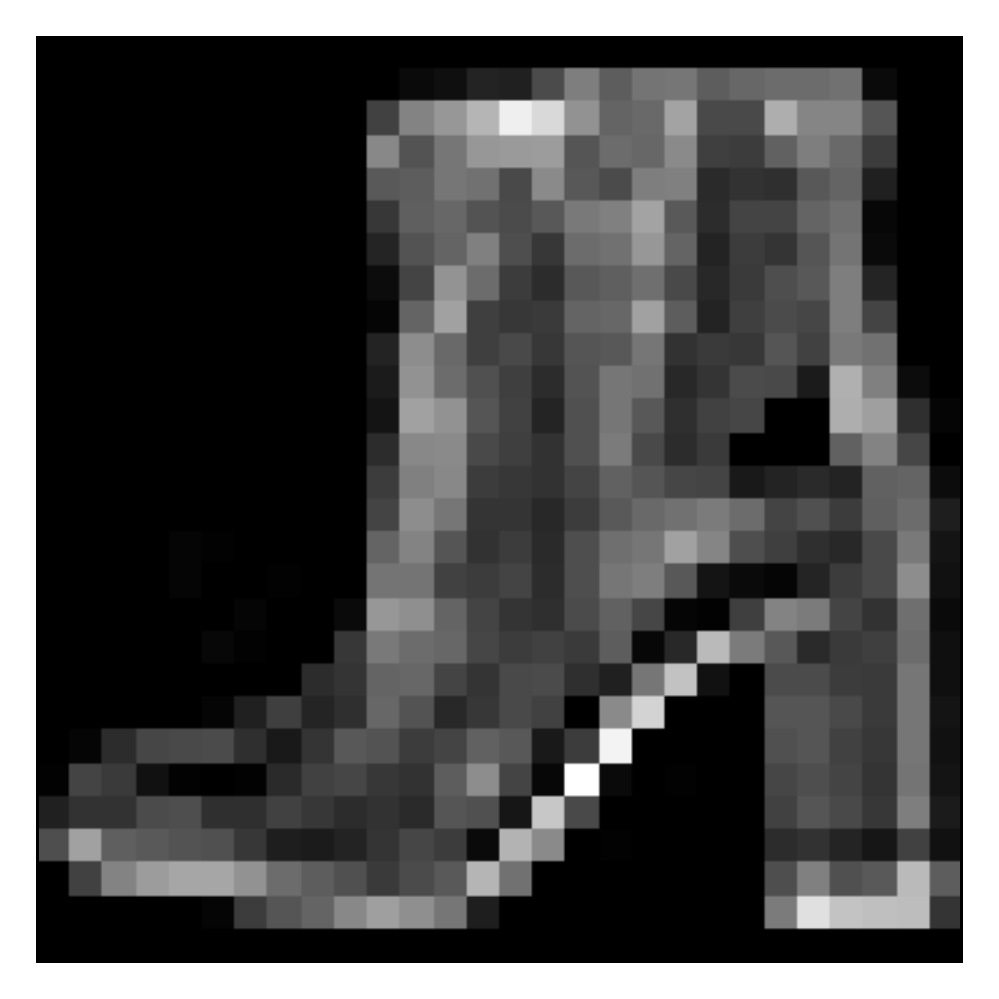}

\end{subfigure}
\begin{subfigure}[b]{0.105\linewidth}
\includegraphics[width=\linewidth]{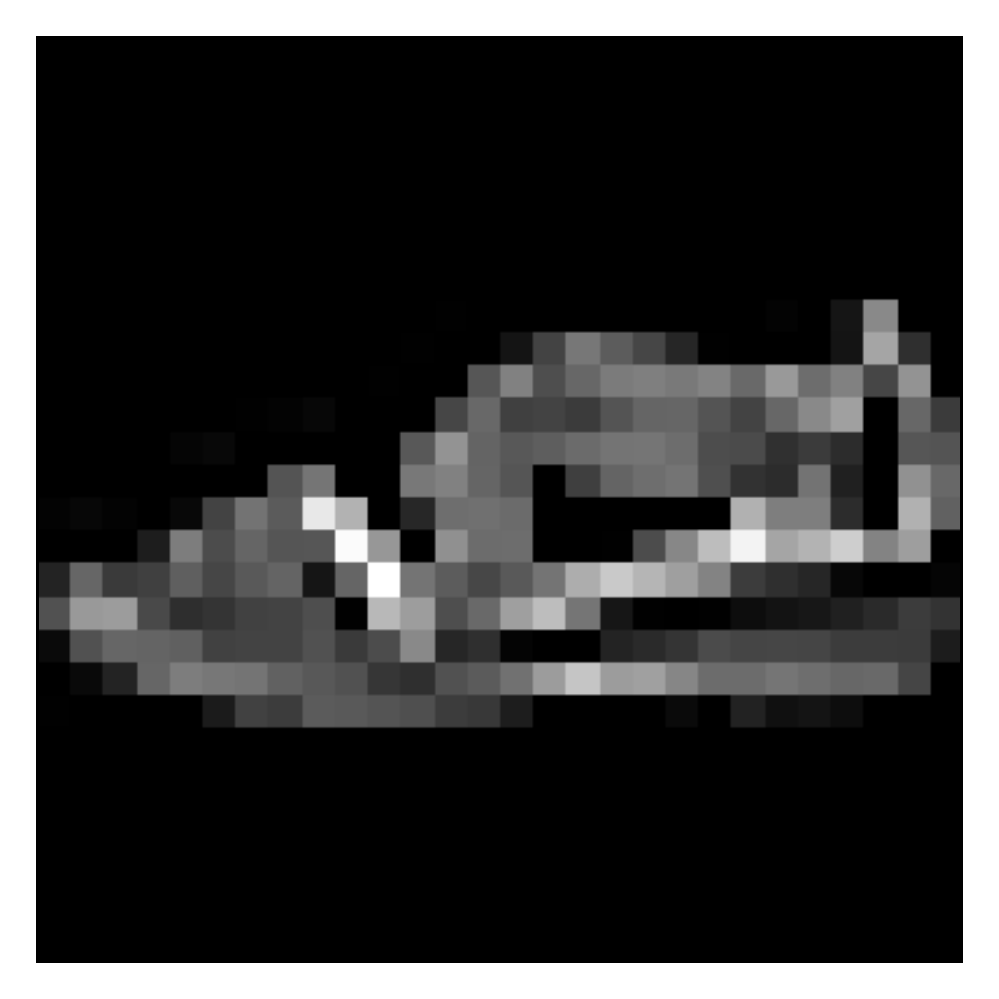}

\end{subfigure}
\begin{subfigure}[b]{0.105\linewidth}
\includegraphics[width=\linewidth]{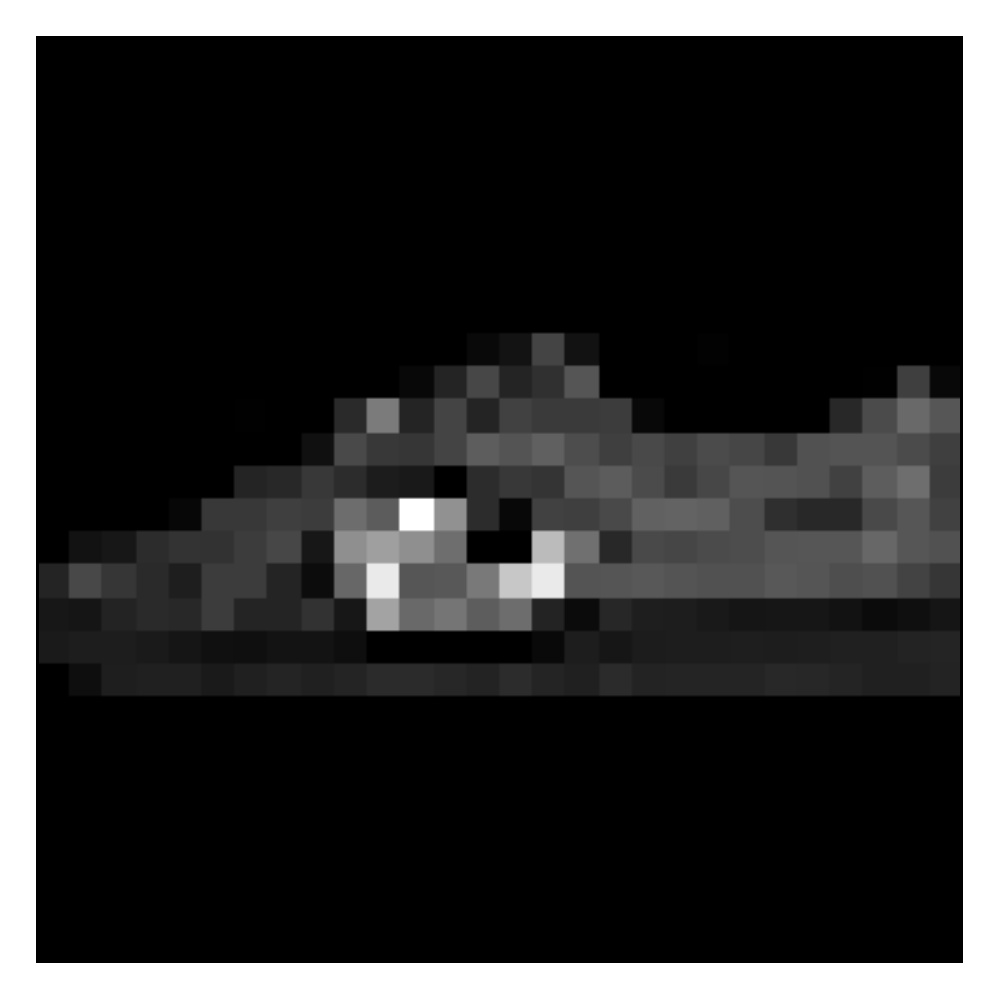}

\end{subfigure}
\begin{subfigure}[b]{0.105\linewidth}
\includegraphics[width=\linewidth]{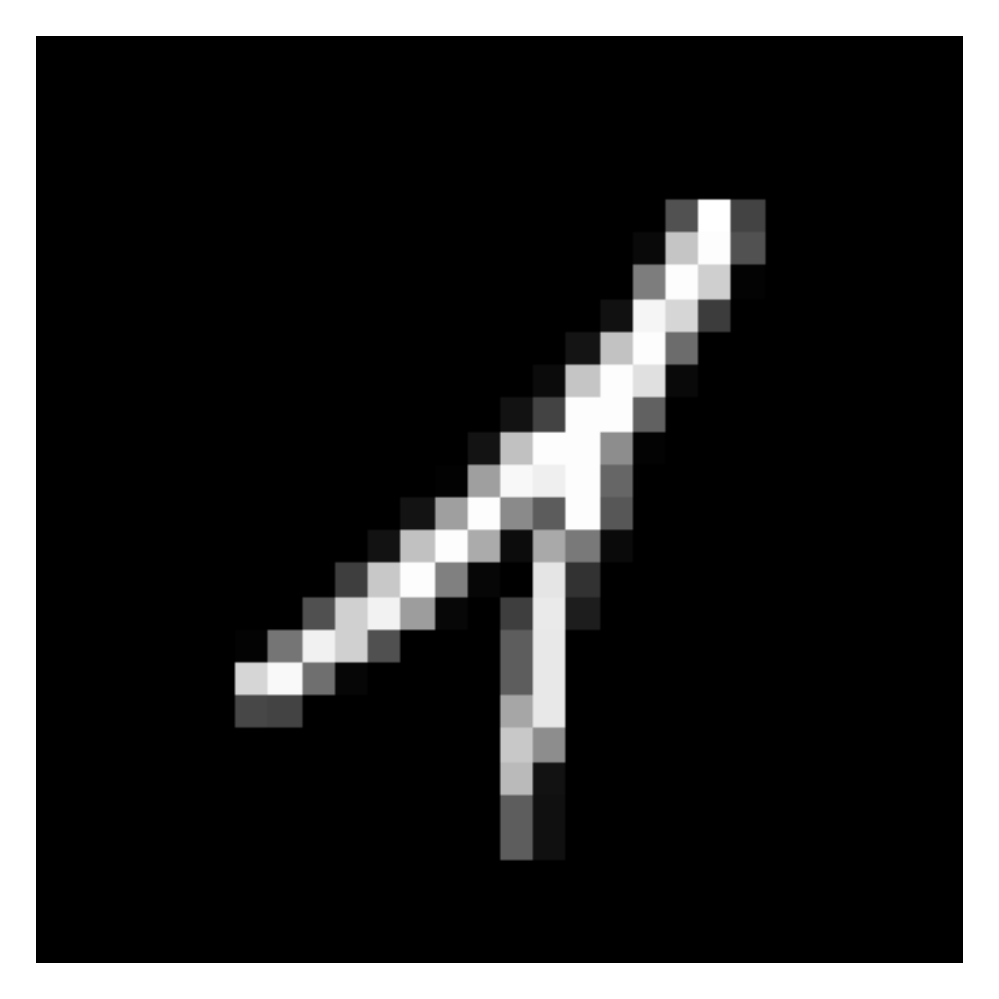}

\end{subfigure}
\begin{subfigure}[b]{0.105\linewidth}
\includegraphics[width=\linewidth]{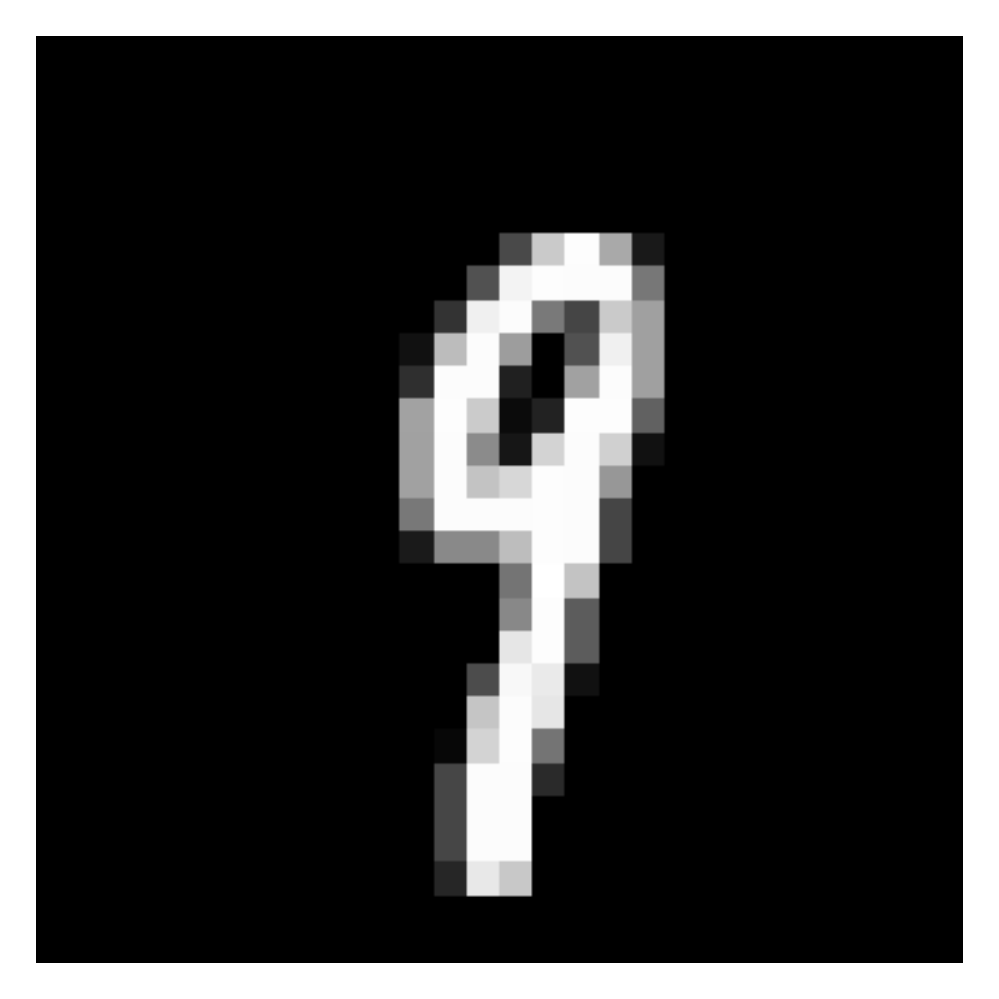}

\end{subfigure}
\begin{subfigure}[b]{0.105\linewidth}
\includegraphics[width=\linewidth]{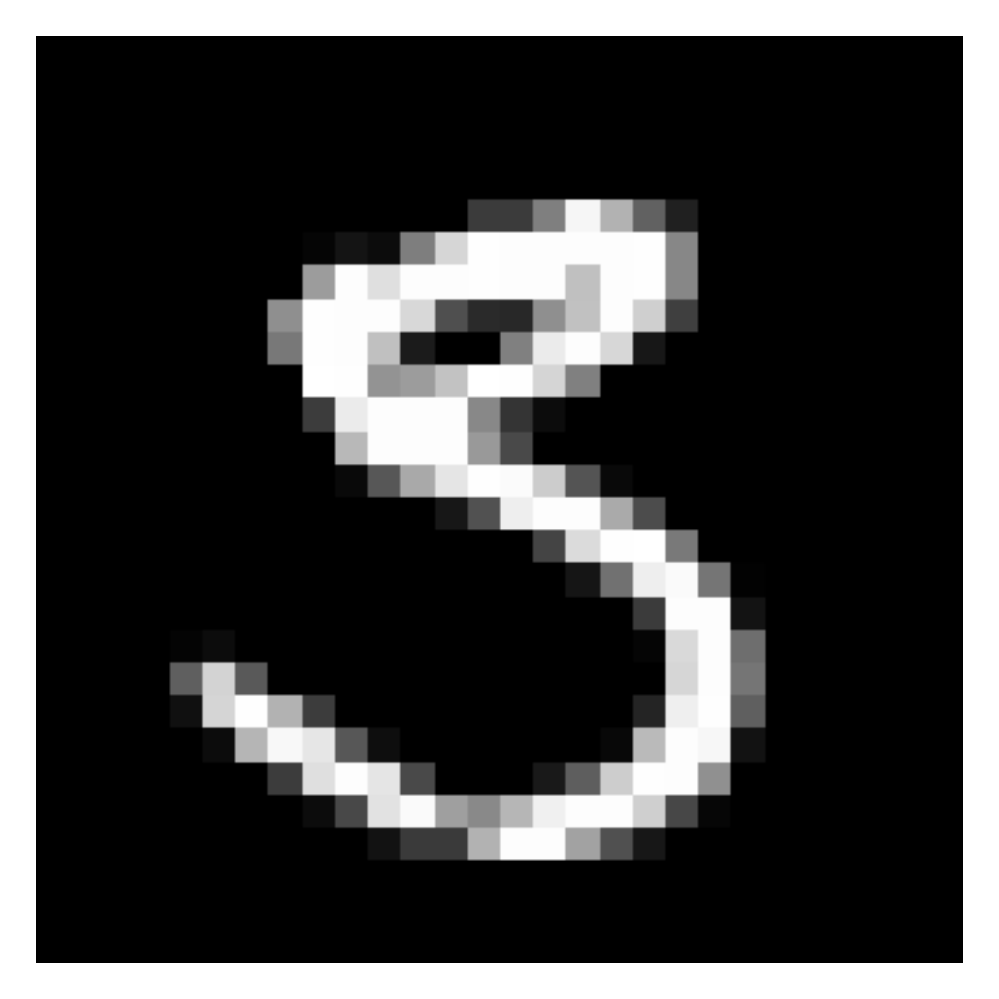}

\end{subfigure}
\begin{subfigure}[b]{0.105\linewidth}
\includegraphics[width=\linewidth]{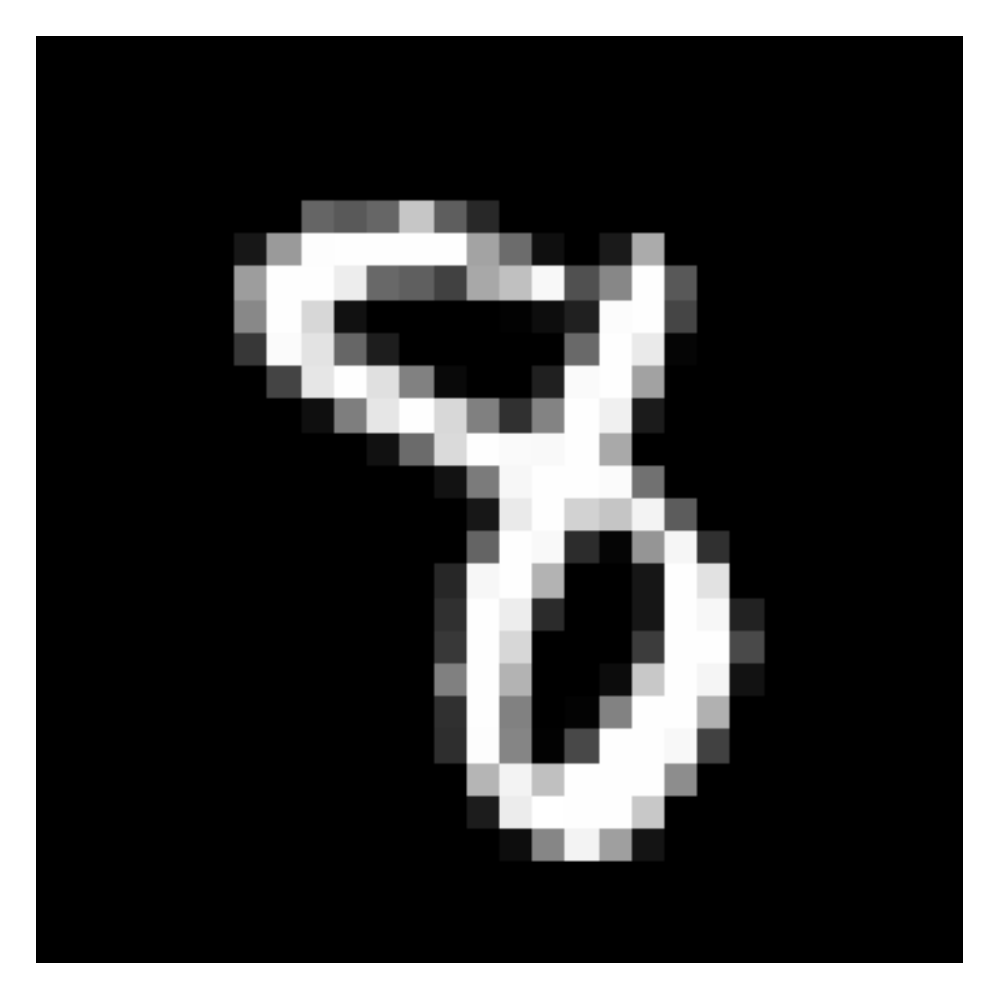}

\end{subfigure}

\begin{subfigure}[b]{0.105\linewidth}
\subcaption*{\textbf{7}: HCLU \newline vs. benign \newline \newline}
\end{subfigure}
\begin{subfigure}[b]{0.105\linewidth}
\includegraphics[width=\linewidth]{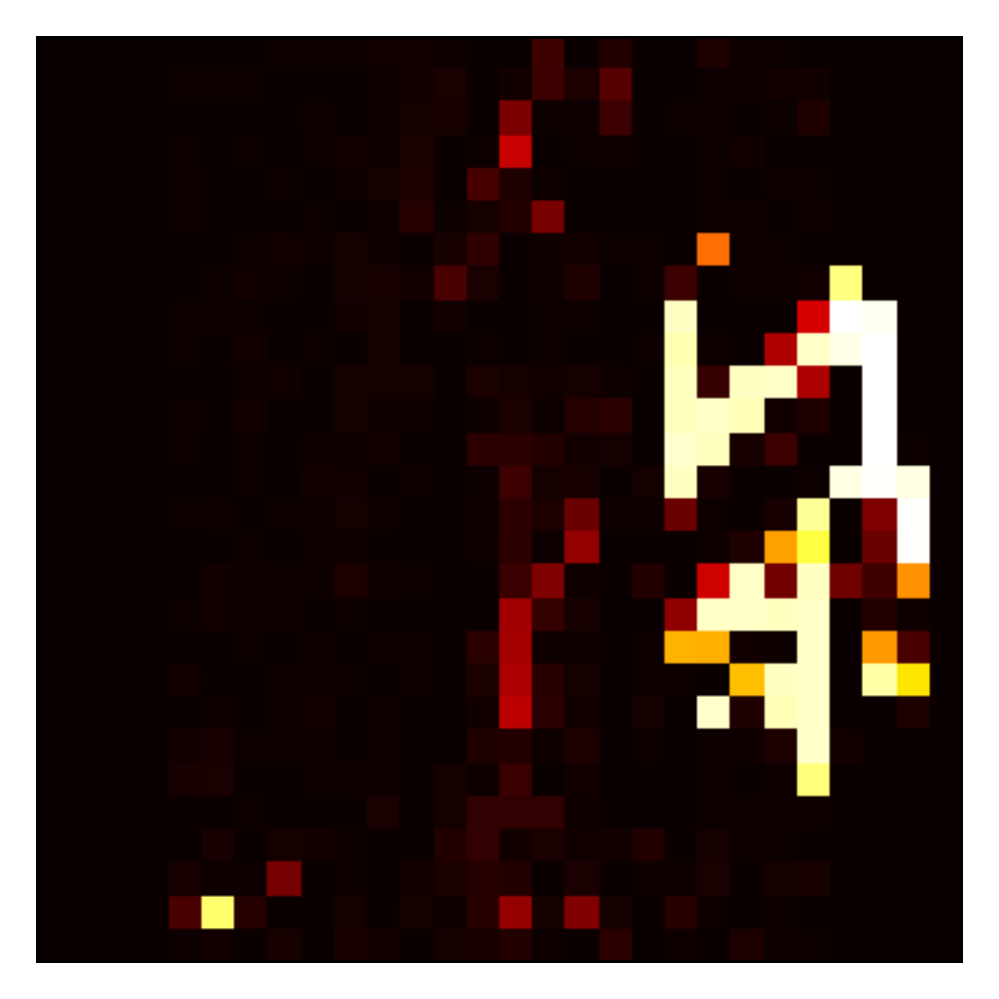}
\subcaption{$\pm .006$.}
\end{subfigure}
\begin{subfigure}[b]{0.105\linewidth}
\includegraphics[width=\linewidth]{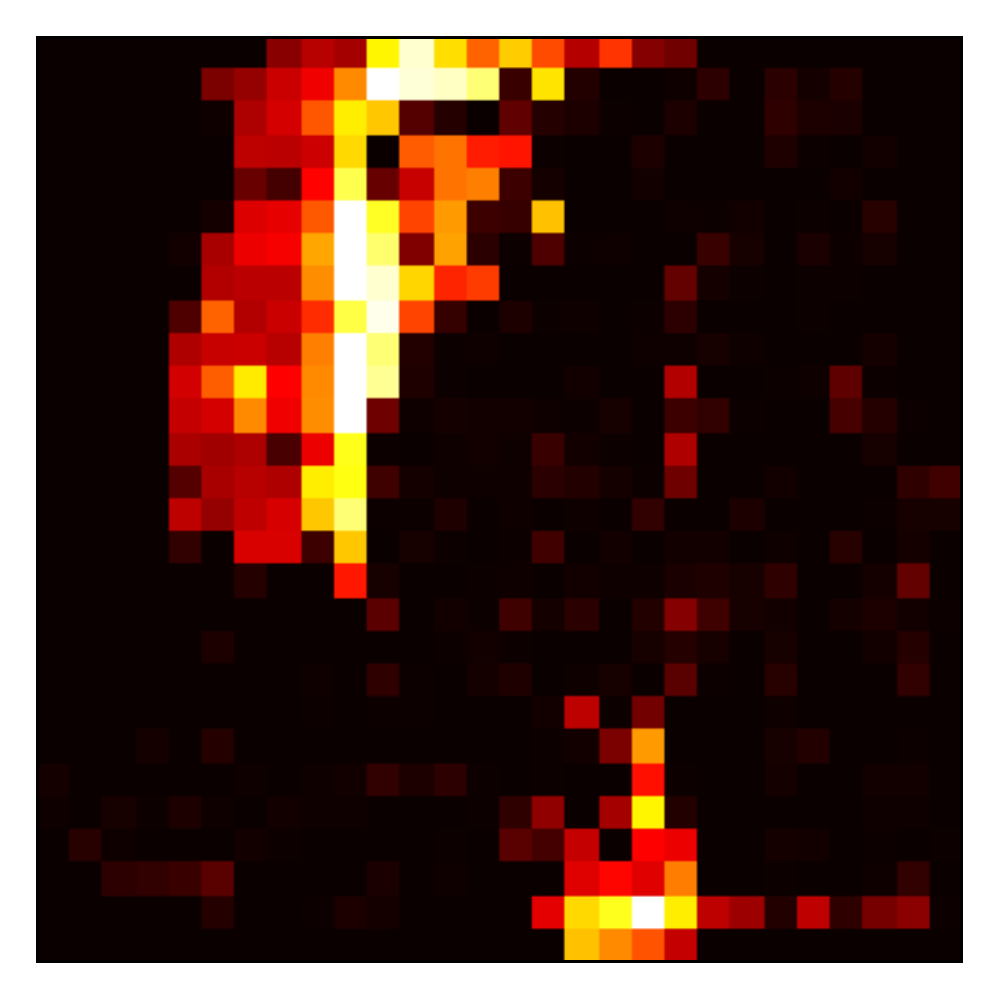}
\subcaption{$\pm .4$.}
\end{subfigure}
\begin{subfigure}[b]{0.105\linewidth}
\includegraphics[width=\linewidth]{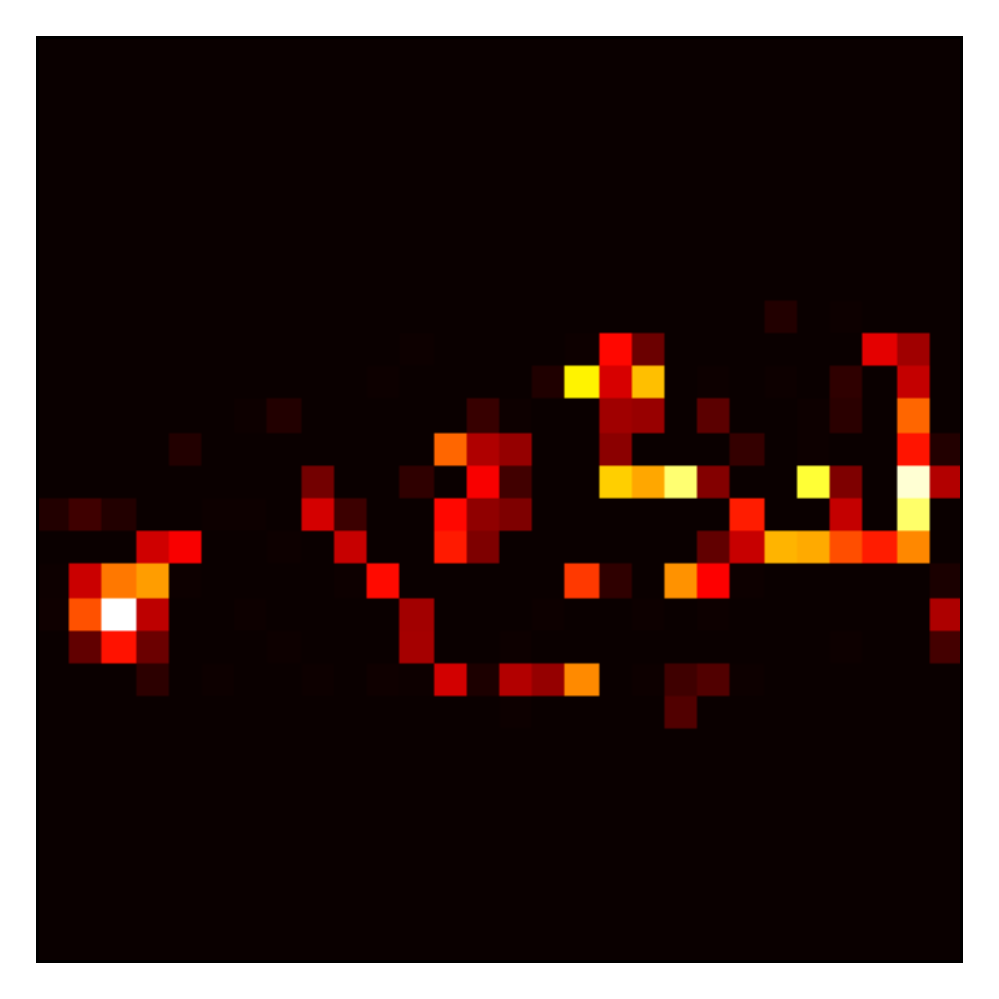}
\subcaption{$\pm 1e^{-6}$.}
\end{subfigure}
\begin{subfigure}[b]{0.105\linewidth}
\includegraphics[width=\linewidth]{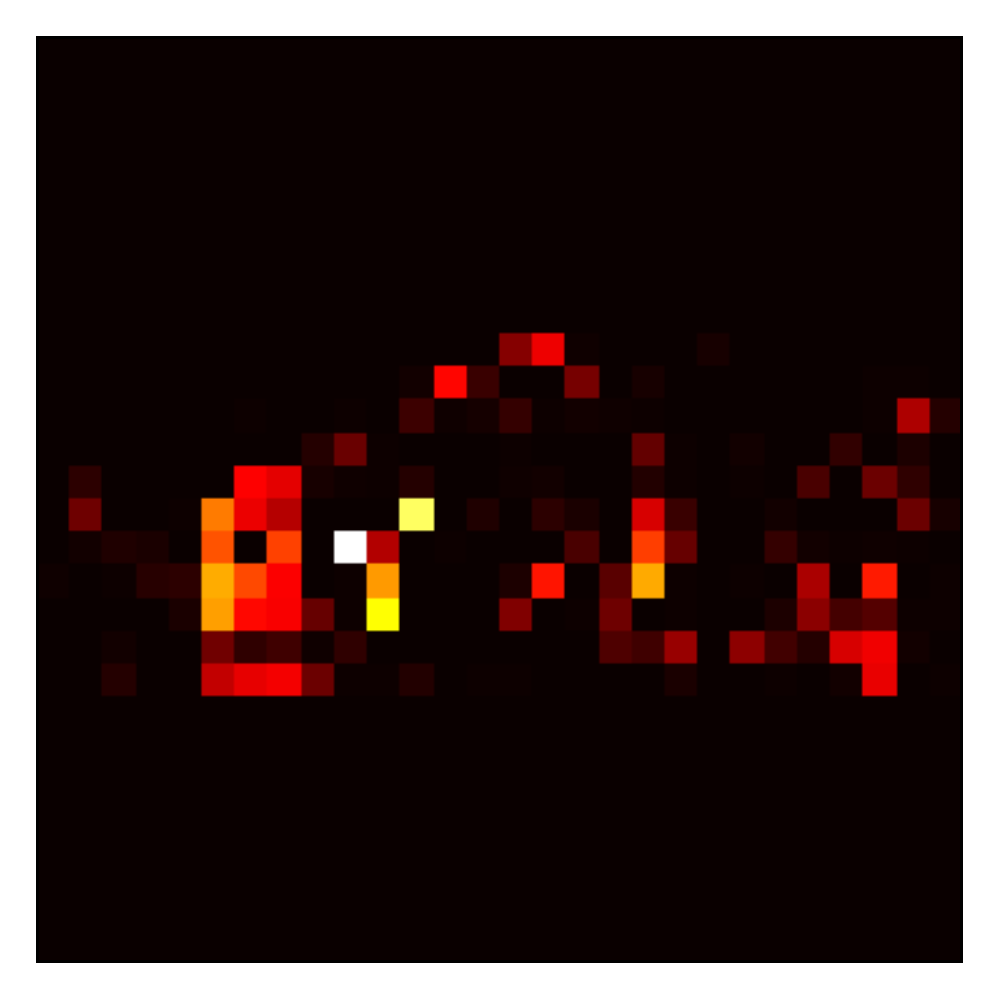}
\subcaption{$\pm1e^{-6}$.}
\end{subfigure}
\begin{subfigure}[b]{0.105\linewidth}
\includegraphics[width=\linewidth]{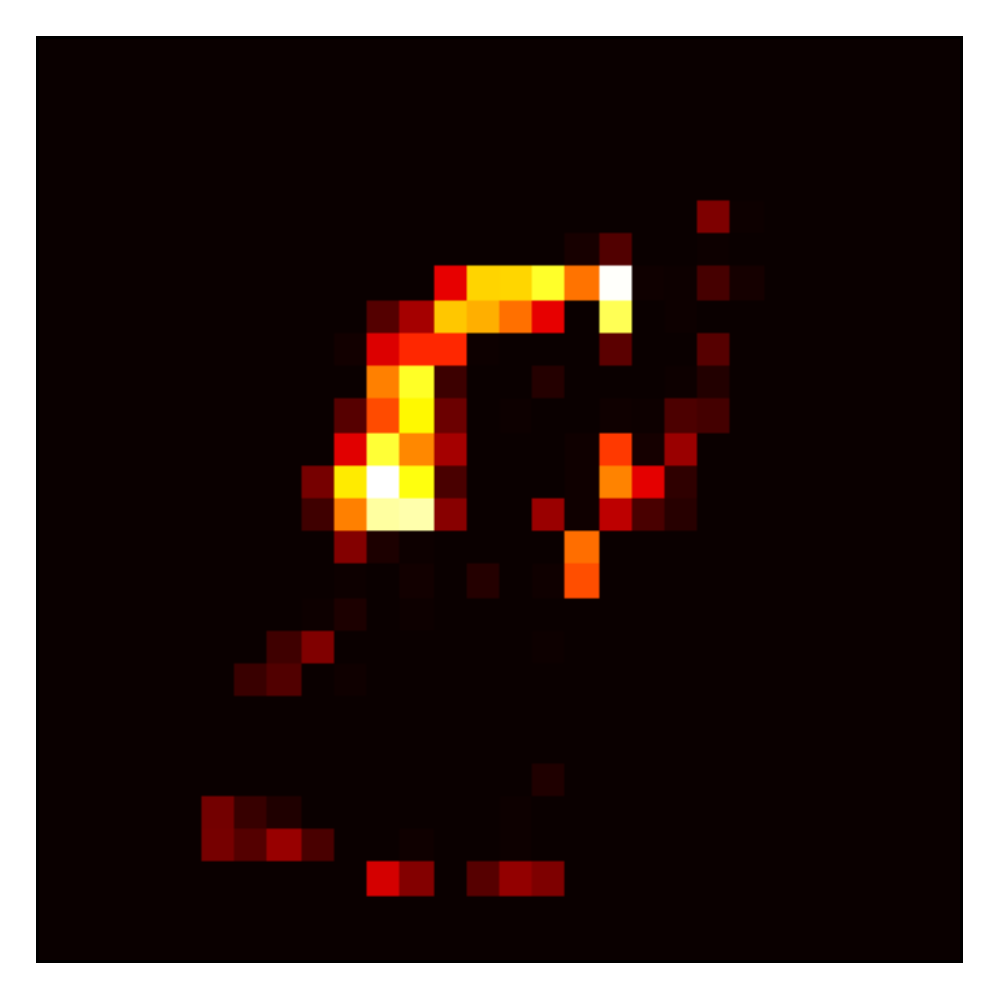}
\subcaption{$\pm .243$.}
\end{subfigure}
\begin{subfigure}[b]{0.105\linewidth}
\includegraphics[width=\linewidth]{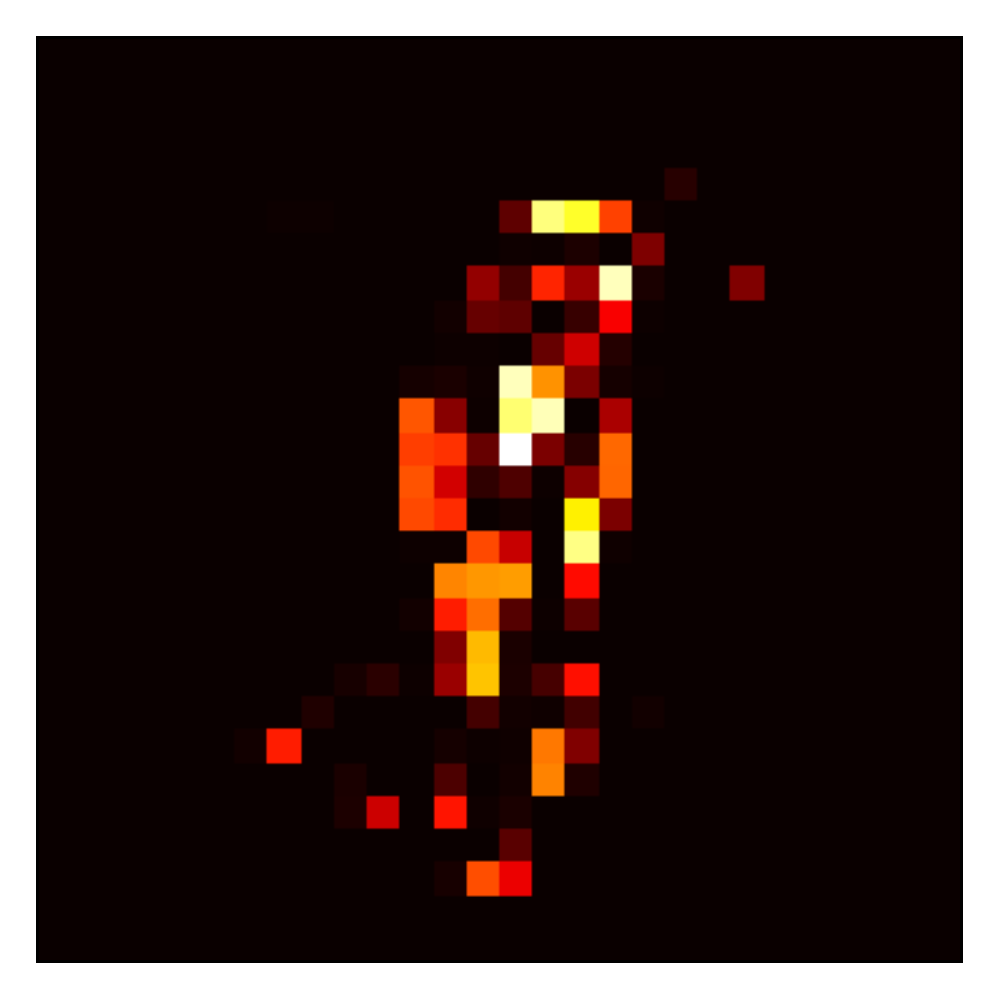}
\subcaption{$\pm 1.0$.}
\end{subfigure}
\begin{subfigure}[b]{0.105\linewidth}
\includegraphics[width=\linewidth]{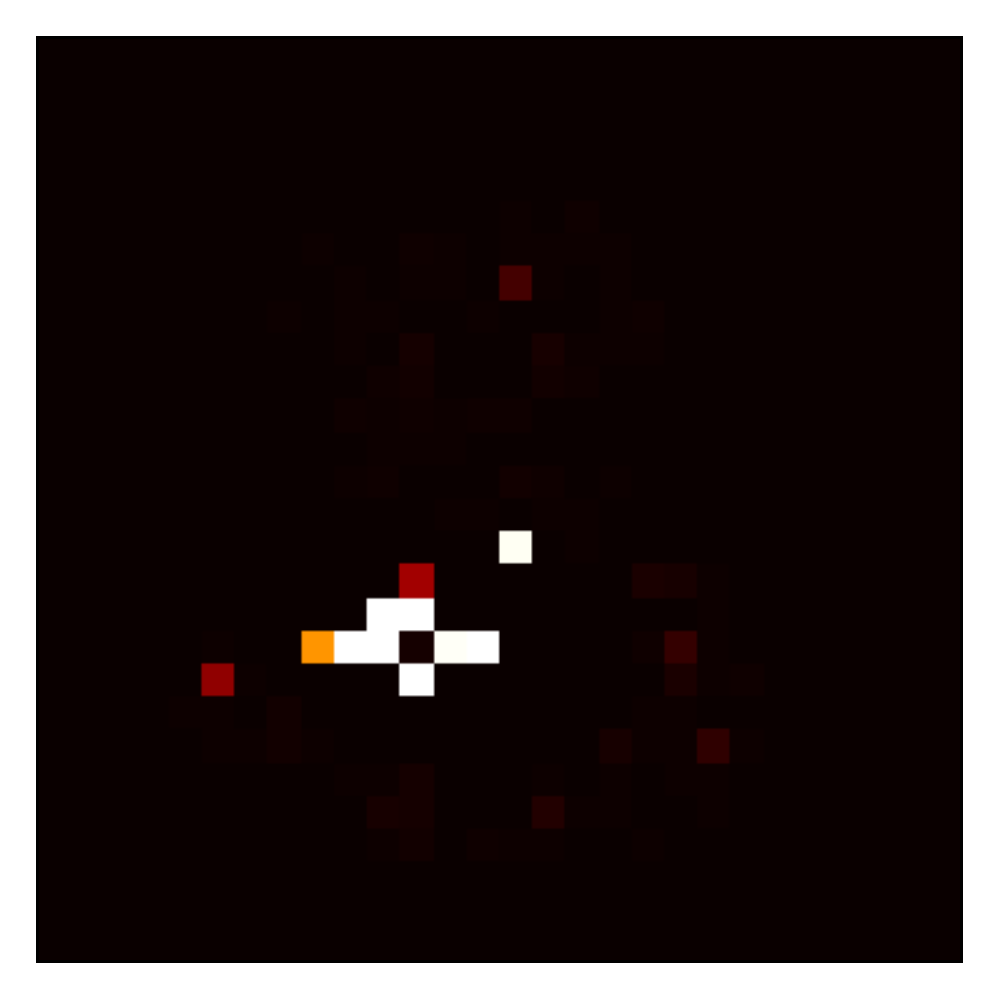}
\subcaption{$\pm .02$.}
\end{subfigure}
\begin{subfigure}[b]{0.105\linewidth}
\includegraphics[width=\linewidth]{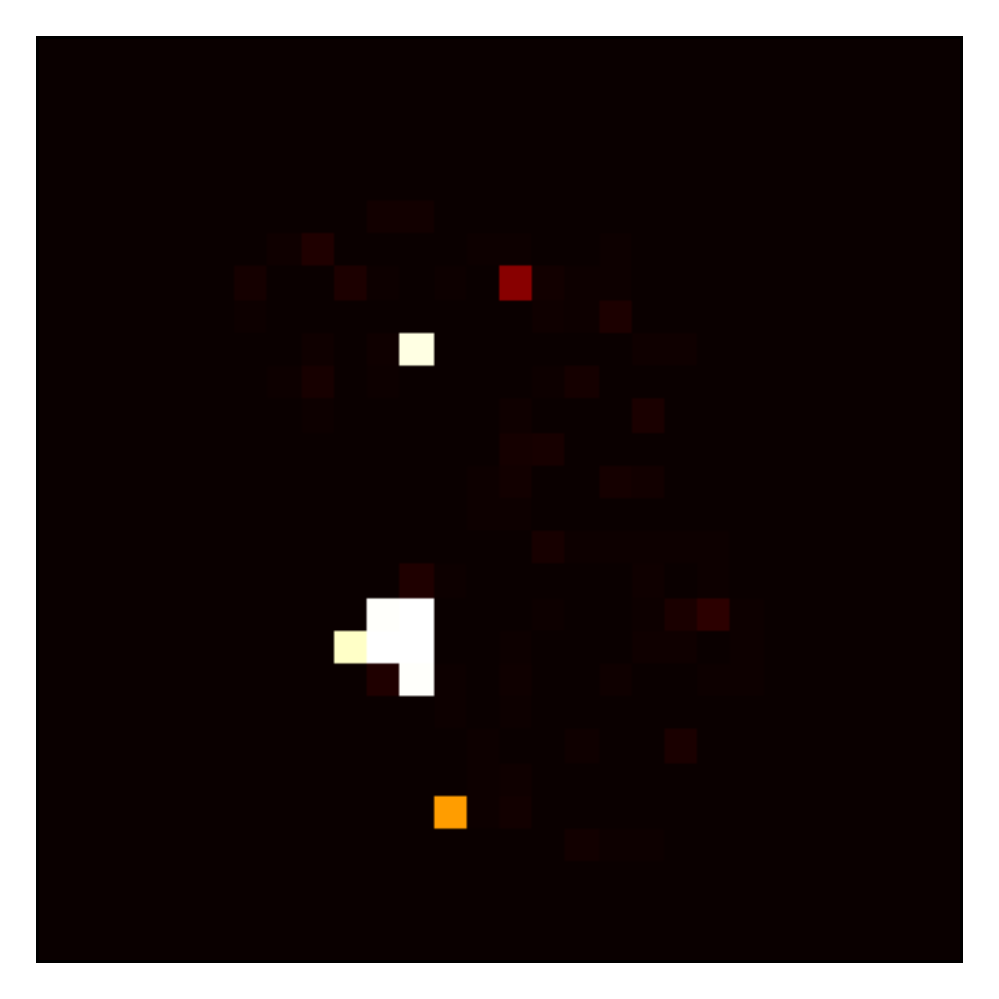}
\subcaption{$\pm 1.0$.}
\end{subfigure}

\caption{Comparison of HCLU, HC, and benign originals. Grey are examples and samples, red plots (row one, four and seven) show differences between images. HC vs. HCLU differences 
 are scaled according to column label. For Fashion MNIST, these scales are also used for the comparisons in row one and seven. We plot all differences using a logarithmic spectrum that allows to see small changes. In the spectrum, colors go from black over red to yellow to white (strongest change). 
Benign originals are the samples started with to craft the upper HCLU  or lower HC example.
 Figures with differences are best seen in color.}
\label{fig:Samples}
\end{figure}

\begin{table}
 \centering 
 \caption{Average size of perturbation and uncertainty change of HC/HCLU to benign data.}\label{table:hclu} 
 \begin{tabular}{@{} l r r r r r@{}} 
 \toprule[1.5pt] 
 & Spam & FMNIST19 & FMNIST57 & MNIST19 & MNIST38 \\ 
  \midrule 

 $\parallel\delta_{HC}\parallel_2$ &0.006$\pm$0.01&0.194$\pm$0.036&0.019$\pm$0.012&0.053$\pm$0.014&0.029$\pm$0.011\\ 
$\parallel\delta_{HCLU}\parallel_2$ &0.008$\pm$0.006&0.194$\pm$0.036&0.019$\pm$0.013&0.053$\pm$0.014&0.03$\pm$0.012\\ 
diff. unc. & 2.35& 0.0& 0.89& 0.31& 0.74\\\bottomrule[1.5pt] 
 \end{tabular} 
\end{table} 

\paragraph{Properties of HCLU examples.}
We show the \hclu \ with the smallest $\delta$ in the fifth row of \cref{fig:Samples}. These examples are still adversarial: we see in the figure that almost all are visually similar to their benign origin. To verify for all depicted examples that they were altered in the crafting process, we plot the original sample beneath the examples.
 The success rates on GPC are 100\%, where however often the specified confidence of $0.95$ is barely not met, and the resulting confidence is around $0.948$. 
\cref{table:hclu} shows the statistics of the adversarial perturbations per feature measures using the $L_2$-norm. 
We observe very small changes on spam, large changes on Fashion MNIST19 and descent changes for all other datasets.  
 
We can also craft examples that only maximize confidence by removing the second constraint.
Surprisingly, the perturbation $\delta_{hc}$ is barely different from the original examples, as visible in \cref{table:hclu}: only on the spam data and two MNIST tasks, a difference is observable. The pictures in the fourth row of \cref{fig:Samples} reveal that albeit looking very similar, the examples are actually different.  
Some features are generally changed, whereas others seem only correlated with uncertainty.
We conclude that slightly different features are learned for confidence and uncertainty, respectively. 
Consequently, for all datasets except Fashion MNIST ankle boot vs trousers, the observed uncertainty is indeed lower when targeted. Concerning the unchanged Fashion MNIST task, we observe that the change in uncertainty for those is examples is only 1\% from the original value. In this case, confidence and uncertainty seem to rely on the same features.

\begin{wrapfigure}{l}{0.49\textwidth}
 \vspace{-10pt}
  \begin{center}
    \includegraphics[width=\linewidth]{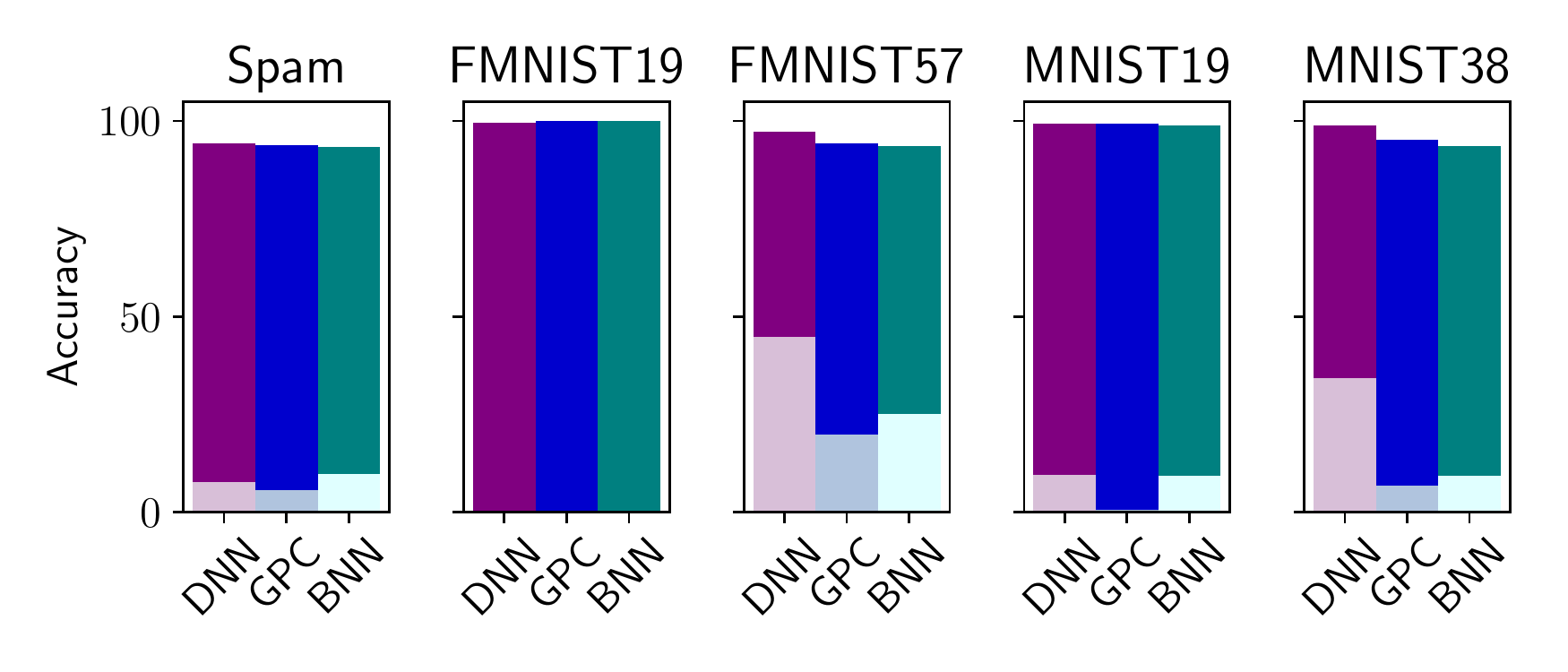}
  \end{center}
\caption{Transferability of optimized GPC adversarial examples. Dark color denotes accuracy on benign data, lighter color on 	\hcluex .}
\label{fig:trans}
\vspace{-5pt}
\end{wrapfigure}

\paragraph{Transferability of misclassification.}
We investigate how many HCLU are misclassified on a GPC (trained on  a different subset of the data), a DNN, and the BNN. The accuracies are 
plotted in \cref{fig:trans}. Dark colors denote benign accuracy, lighter colors the accuracy on HCLU.

The decrease in accuracy is similar on all datasets, 
except the Fashion MNIST sandals vs sneakers and the DNN on MNIST three vs eight. 
In two cases, we observe that the DNN shows slightly higher accuracy 
on HCLU: on the MNIST data and Fashion MNIST sandals vs sneakers. Still, 
in all cases, accuracy is reduced significantly for HCLU examples.

\paragraph{Transferability of uncertainty.}
We now test the effect of \hclu \ on Bayesian neural networks (BNN) uncertainty measures. For misclassification, e.g. non-Bayesian decision boundaries, \cite{DBLP:journals/corr/PapernotMG16} showed that adversarial examples often fool several models. 
In this experiment, we are interested whether behavior 
\emph{differs} between benign and adversarial data. 
We chose Carlini and Wagner's $L_2$ attack as a baseline: $L_2$, as in our case, allows the best optimization.  We further configure the attack as to increase transferability of the examples (corresponding to higher ``confidence'' on the target DNN)\footnote{We set $\kappa=0.7$. The attack definition, in a nutshell, defines $\kappa>0$ to encourage the solver to find a confidently classified example. For details see \cite{DBLP:journals/corr/CarliniW16a} .}.

\begin{wrapfigure}{r}{0.55\textwidth}
\vspace{-10pt}
\includegraphics[width=\linewidth]{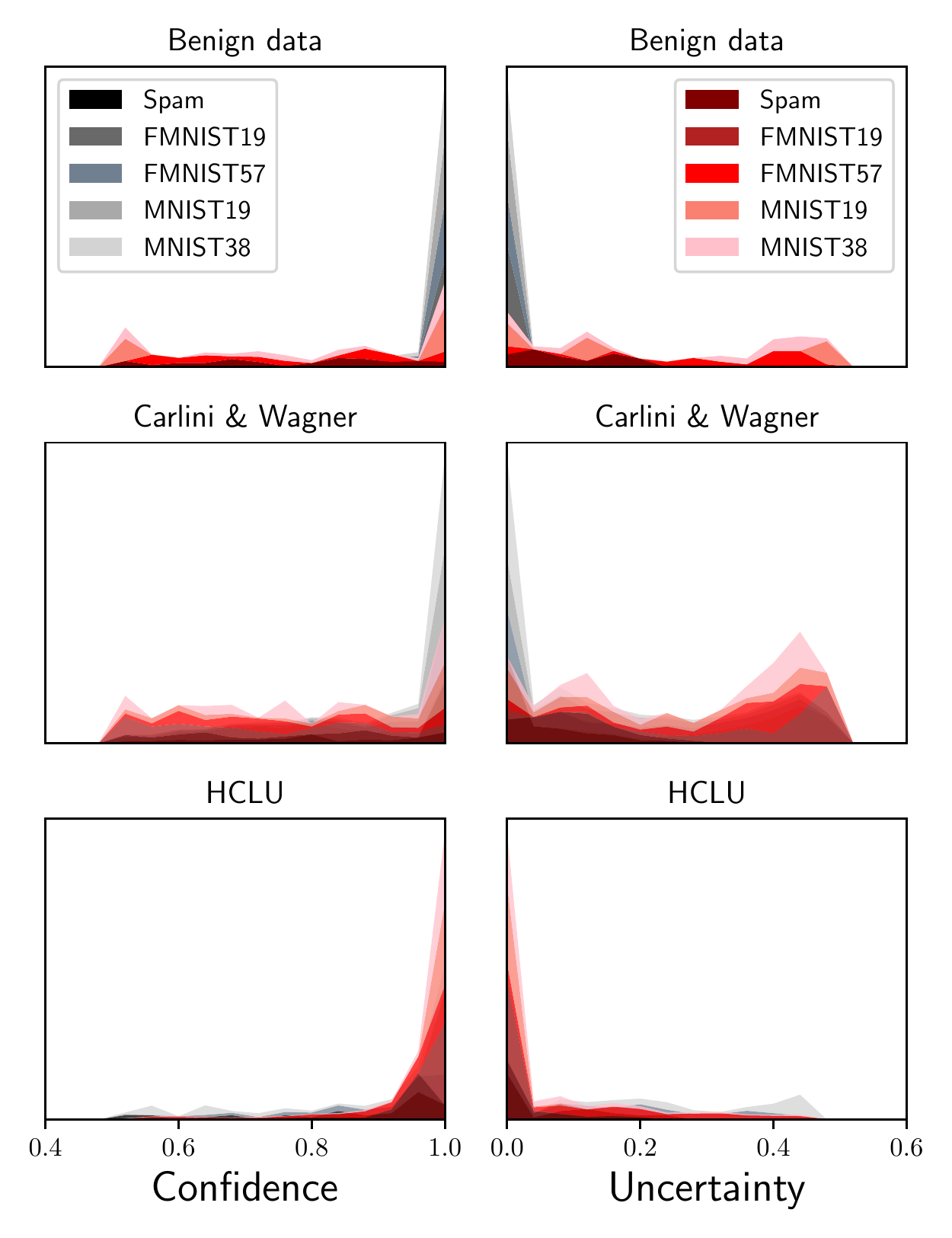}

\caption{Transferability of HCLU examples (bottom) to Bayesian Neural Networks. We consider Carlini \& Wagner's $L_2$ attack as a comparison (middle). Benign data is also depicted as a baseline (top). Correctly classified data is plotted in gray shades, misclassified data in red shades. Figure is best seen in color.}
\label{fig:BNN_unc}
\end{wrapfigure}

The accuracy on these $L_2$ adversarial examples on DNN, GPC, and BNN is higher than on HCLU: the average accuracy is between $50$\% and in some cases higher than $90$\%. Yet, the accuracy under the transferred attack is always lower than the accuracy observed using clean test data.

We depict the results concerning Bayesian confidence and uncertainty in \cref{fig:BNN_unc}, where we distinguish correctly classified (gray shades) and wrongly classified (red shades) benign and adversarial data. We measure the mean (confidence, left plots) and variance (uncertainty, right plots) of the sampled posteriors and bin them using $25$ bins between $0.0$ and $1.0$. As large factions of the histograms are empty, we plot only the relevant parts. To outline overall trends, we plot the normalized bins of correct and wrongly classified data stacked on top of each other. 

In general, the BNN is more confident on benign data/Carlini and Wagner examples that are correctly classified.
This observation holds across all data sets. 
For HCLU, this trend is reversed:
the BNN is confident on many misclassified examples. 
Intriguingly, the BNN outputs low confidence on some \hcluex \ which are not misclassified, or correctly assigned to their original class.  
Analogously, the uncertainty measures are similar between benign data and Carlini and Wagner's attack. Uncertainty is generally low for correctly classified data and high for wrongly classified data. This observations are again reversed for HCLU examples: here uncertainty is often low if an example is wrongly classified.

\section{Conclusion}
\label{sec:conclusion} 
In this paper, we studied 
the vulnerability of machine learning models providing Bayesian model uncertainty.
 introduced a technique to craft HCLU adversarial examples, which achieve both high confidence and low uncertainty on a Gaussian process classifier. 

Also Bayesian neural networks misclassify HCLU. We further found that HCLU adversarial examples are misclassified with low uncertainty and high confidence, in contrast to high-confidence adversarial examples which are derived from non-Bayesian models such as DNN. 

Our work has several implications. Firstly, different Bayesian models (GPC, BNN) learn similar confidence and uncertainty features, as our transferability study shows. This implies that such measures cannot be used as a defense. Our results show as well that there is a subtle difference in the features learned for uncertainty and confidence, respectively.  

\section*{Acknowledgments}
This work was supported by the German Federal Ministry of Education and
Research (BMBF) through funding for the Center for IT-Security,
Privacy and Accountability (CISPA) (FKZ: 16KIS0753). This work has further been supported by the Engineering and Physical Research Council (EPSRC) Research Project EP/N014162/1.

\bibliographystyle{unsrtnat}

\bibliography{advgp}

\end{document}